\documentclass{article}

\usepackage{fullpage}
\usepackage{textcomp}
\usepackage{setspace}
\usepackage{enumerate}
\usepackage{amsmath}
\usepackage{amsfonts}
\usepackage{amssymb}
\usepackage{graphicx}
\usepackage{mathrsfs}
\usepackage{geometry}
\usepackage{caption}
\usepackage{subcaption}

\newtheorem{lemma}{Lemma}
\numberwithin{lemma}{section} % important bit
\newtheorem{fact}[lemma]{Fact}
\newtheorem{theorem}[lemma]{Theorem}
\newtheorem{corollary}[lemma]{Corollary}
\newtheorem{definition}[lemma]{Definition}

\title{\LARGE Searching for the closest-pair in a query translate}
\bigskip

\author{
  Jie Xue\footnote{University of Minnesota - Twin Cities, MN, USA; \texttt{\{xuexx193,janardan\}@umn.edu}.}
  \and
  Yuan Li\footnote{Facebook Inc., Seattle, WA, USA; \texttt{lydxlx@fb.com}.}
  \and
  Saladi Rahul\footnote{University of Illinois at Urbana-Champaign, IL, USA; \texttt{saladi.rahul@gmail.com}.}
  \and
  Ravi Janardan\footnotemark[1]
}
\date{}

\begin{document}

\maketitle

%\pagenumbering{gobble}% Remove page numbers (and reset to 1)
\begin{abstract}
We consider a range-search variant of the closest-pair problem.
Let $\varGamma$ be a fixed shape in the plane.
We are interested in storing a given set of $n$ points in the plane in some data structure such that for any specified translate of $\varGamma$, the closest pair of points contained in the translate can be reported efficiently.
We present results on this problem for two important settings: when $\varGamma$ is a polygon (possibly with holes) and when $\varGamma$ is a general convex body whose boundary is smooth.
When $\varGamma$ is a polygon, we present a data structure using $O(n)$ space and $O(\log n)$ query time, which is asymptotically optimal.
When $\varGamma$ is a general convex body with a smooth boundary, we give a near-optimal data structure using $O(n \log n)$ space and $O(\log^2 n)$ query time.
Our results settle some open questions posed by Xue et al. [SoCG 2018].
\end{abstract}

%\newpage
%\clearpage
%\pagenumbering{arabic}% Arabic page numbers (and reset to 1)

\section{Introduction}
The range closest-pair (RCP) problem, as a range-search version of the closest-pair problem, aims to store a given set $S$ of $n$ points in some data structure such that for a specified query range $X \in \mathcal{X}$ chosen from a certain \textit{query space} $\mathcal{X}$, the closest pair of points in $S \cap X$ can be reported efficiently.
As a range-search problem, the RCP problem is non-decomposable in the sense that even if the query range $X$ can be written as $X = X_1 \cup X_2$, the closest-pair in $S \cap X$ cannot be determined efficiently knowing the closest-pairs in $S \cap X_1$ and $S \cap X_2$.
The non-decomposability makes the problem quite challenging and interesting, as many traditional range-search techniques are inapplicable.

The RCP problem in $\mathbb{R}^2$ has been well-studied over years \cite{abam2009power,bae2018closest,gupta2006range,gupta2014data,shan2003spatial,sharathkumar2007range,xue2019colored,xue2018approximate,xue2018new}.
%and data structures have been designed for various query spaces including quadrants, strips, rectangles, and halfplanes
Despite of much effort, the query ranges considered are still restricted to very simple shapes, typically orthogonal rectangles and halfplanes.
It is then interesting to ask what if the query ranges are of more general shapes.
%In this paper, we try to extend the scope of the RCP problem by considering query ranges of more general shapes.
%If there is no restriction on the query ranges at all, the problem seems unsolvable\footnote{No solution is known even for classical range-search problems.}.
In this paper, we consider a new variant of the RCP problem in which the query ranges are \textit{translates} of a fixed shape (which can be quite general).
Formally, let $\varGamma$ be a fixed shape in $\mathbb{R}^2$ called \textit{base shape} and $\mathcal{L}_\varGamma$ be the collection of all translates of $\varGamma$.
We investigate the RCP problem with the query space $\mathcal{L}_\varGamma$ (or the $\mathcal{L}_\varGamma$-RCP problem).
This type of query, which is for the first time mentioned in \cite{xue2018new} (as an open question), is natural and well-motivated.
First, in range-search problems, the query spaces considered are usually closed under translation; in this sense, the query space consisting of translates of a single shape seems the most ``fundamental'' query type.
Some of the previously studied query ranges, e.g., quadrants and halfplanes \cite{abam2009power,gupta2014data,xue2018new}, are in fact instances of translation queries (halfplanes can be viewed as translates of an ``infinitely'' large disc).
Also, translation queries find motivation in practice.
For instance, in many applications, the user may be interested in the information within a certain distance $r$ from him/her.
In this situation, the query ranges are discs of a fixed radius $r$, i.e., translates of a fixed disc; or more generally, if the distance $r$ is considered under a general distance function induced by a norm $\lVert \cdot \rVert$, then the query ranges are translates of a $\lVert \cdot \rVert$-disc of radius $r$.
Finally, there is another view of the translation queries: the base shape $\varGamma$ can be viewed as static while the dataset is translating.
With this view, a motivation of the translation queries is to monitor the information in a fixed region (i.e., $\varGamma$) for moving points (where the movement pattern only includes translation).

%The problem appears to be quite challenging if the base shape $\varGamma$ is (totally) arbitrary.
%Thus, it is reasonable to make assumptions about $\varGamma$.
We investigate the problem in two important settings: when $\varGamma$ is a polygon (possibly with holes) and when $\varGamma$ is a general convex body whose boundary is smooth (i.e., through each point on the boundary there is a unique tangent line to $\varGamma$).
Our main goal is to design optimal or near-optimal data structures for the problems in terms of \textit{space cost} and \textit{query time}.
The preprocessing of these data structures is left as an open question for future study.

Although we restrict the query ranges to be translates of a fixed shape, the problem is still challenging for a couple of reasons.
%First, the RCP problem is a non-decomposable range-search problem in the sense that even if the query range $X$ can be written as $X = X_1 \cup X_2$, the closest-pair in $S \cap X$ cannot be determined efficiently given the closest-pairs in $S \cap X_1$ and $S \cap X_2$.
%The non-decomposability makes many traditional range-search techniques inapplicable to the RCP problem.
First, the base shape $\varGamma$ to be considered is quite general in both of our settings.
When $\varGamma$ is a polygon, it needs not be convex, and indeed can even have holes.
In the case where $\varGamma$ is a general convex body, we only need the aforementioned smoothness of its boundary.
Second, we want the RCP data structures to be optimal or near-optimal, namely, use $O(n \cdot \text{poly}(\log n))$ space and have $O(\text{poly}(\log n))$ query time.
This is usually difficult for a non-decomposable range-search problem.

\subsection{Related work and our contributions}
\textbf{Related work.}
The closest-pair problem and range search are both classical topics; some surveys can be found in \cite{agarwal1999geometric:range_search,smid2000closest}.
The RCP problem in $\mathbb{R}^2$ has been studied in prior work \cite{abam2009power,bae2018closest,gupta2006range,gupta2014data,shan2003spatial,sharathkumar2007range,xue2019colored,xue2018approximate,xue2018new}.
%The query types considered include quadrants, strips, rectangles, and halfplanes.
State-of-the-art RCP data structures for quadrant, strip, rectangle, and halfplane queries were given in the recent work \cite{xue2018new}.
The quadrant and halfplane RCP data structures are optimal (i.e., with linear space and logarithmic query time).
The strip RCP data structure uses $O(n \log n)$ space and $O(\log n)$ query time, while the rectangle RCP data structure uses $O(n \log^2 n)$ space and $O(\log^2 n)$ query time.
The work \cite{xue2019colored} considered a colored version of the RCP problem and gave efficient approximate data structures.
The paper \cite{xue2018approximate} studied an approximate version of the RCP problem in which the returned answer can be slightly outside the query range.
\medskip

\noindent
\textbf{Our contributions.}
We investigate a new variant of the RCP problem in which the query ranges are translates of a fixed shape $\varGamma$.
%We study the problem when $\varGamma$ is a polygon (possibly with holes) and when $\varGamma$ is a general convex body with a smooth boundary.
In the first half of the paper, we assume $\varGamma$ is a fixed polygon (possibly with holes), and give an RCP data structure for $\varGamma$-translation queries using $O(n)$ space and $O(\log n)$ query time, which is asymptotically optimal.
In the second half of the paper, we assume $\varGamma$ is a general convex body with a smooth boundary, and give a near-optimal RCP data structure for $\varGamma$-translation queries using $O(n \log n)$ space and $O(\log^2 n)$ query time.
The $O(\cdot)$ above hides constants depending on $\varGamma$.
Our results settle some open questions posed in \cite{xue2018new}, e.g., the RCP problem with fixed-radius disc queries, etc.
In order to design these data structures, we make nontrivial geometric observations and exploit the properties of the problem itself (i.e., we are searching for the \textit{closest-pair} in a \textit{translate}).
Many of our intermediate results are of independent interest and can probably be applied to other related problems.
We describe our key ideas and techniques in Section~\ref{sec-technique} after establishing relevant notations in Section~\ref{sec-pre}.
\smallskip

\noindent
\textbf{Organization.}
Section~\ref{sec-pre} presents the notations and preliminaries used throughout the paper.
%We suggest that the readers read this section carefully before moving on.
Section~\ref{sec-technique} gives an overview of the techniques we use to solve the problems.
In Section~\ref{sec-polygon}, we study the problem when $\varGamma$ is a polygon.
In Section~\ref{sec-general}, we study the problem when $\varGamma$ is a general convex body with a smooth boundary.
To make the paper more readable, some proofs and details are deferred to the appendix.

\subsection{Preliminaries} \label{sec-pre}
\textbf{Basic notations and concepts.}
For $a,b \in \mathbb{R}^2$, we use $\text{dist}(a,b)$ to denote the Euclidean distance between $a$ and $b$, and use $[a,b]$ to denote the segment connecting $a$ and $b$.
The \textit{length} of a pair $\phi = (a,b)$ of points, denoted by $|\phi|$, is the length of the segment $[a,b]$, i.e., $|\phi| = \text{dist}(a,b)$.
For a shape $\varGamma$ in $\mathbb{R}^2$ and a point $p \in \mathbb{R}^2$, we denote by $\varGamma_p$ the $\varGamma$-translate $p+\varGamma$.
We write $\mathcal{L}_\varGamma = \{\varGamma_p: p \in \mathbb{R}^2\}$, i.e., the collection of all $\varGamma$-translates.
\smallskip

\noindent
\textbf{Candidate pairs.}
%The \textit{length} of a pair $(a,b)$ of points refers to the length of the segment $[a,b]$.
Let $S$ be a set of points in $\mathbb{R}^2$ and $\mathcal{X}$ a collection of ranges.
A \textit{candidate pair} in $S$ with respect to $\mathcal{X}$ refers to a pair of points in $S$ that is the closest-pair in $S \cap X$ for some $X \in \mathcal{X}$.
We denote by $\varPhi(S,\mathcal{X})$ the set of the candidate pairs in $S$ w.r.t. $\mathcal{X}$.
\smallskip

\begin{figure}[h]
  \begin{center}
    \includegraphics[height=2.5cm]{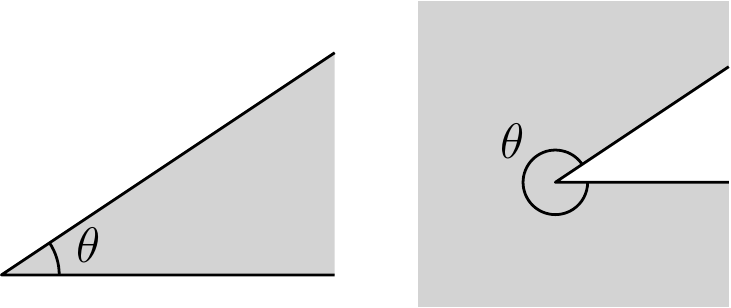}
  \end{center}
  \caption{Examples of wedges and co-wedges.}
  \label{fig-weco}
\end{figure}
\noindent
\textbf{Wedges and co-wedges.}
A \textit{wedge} is a range in $\mathbb{R}^2$ defined by an angle $\theta \in (0,\pi)$, which is the intersection of two halfplanes (see the left figure in Figure~\ref{fig-weco}).
A \textit{co-wedge} is a range in $\mathbb{R}^2$ defined by an angle $\theta \in (\pi,2\pi)$, which is the union of two halfplanes (see the right figure in Figure~\ref{fig-weco}).
The boundary of a wedge or co-wedge $W$ consists of two rays sharing a common initial point, called the two \textit{branches} of $W$.
When appropriate, we refer to wedges and co-wedges collectively as (\textit{co}-)\textit{wedges}.
\smallskip

\noindent
\textbf{Convex bodies.}
A \textit{convex body} in $\mathbb{R}^2$ refers to a compact convex shape with a nonempty interior.
If $C$ is a convex body in $\mathbb{R}^2$, we denote by $\partial C$ the boundary of $C$, which is a simple cycle, and by $C^\circ$ the interior of $C$, i.e., $C^\circ = C \backslash \partial C$.
\smallskip

\noindent
The following two lemmas will be used in various places in this paper.
\begin{lemma} \label{lem-pigeon}
	Let $\varGamma$ be a fixed bounded shape in $\mathbb{R}^2$, and $\mu>0$ be a constant.
	Also, let $S$ be a set of points in $\mathbb{R}^2$.
	Then for any point $p \in \mathbb{R}^2$, either the closest-pair in $S \cap \varGamma_p$ has length smaller than $\mu$, or $|S \cap \varGamma_p| = O(1)$.
\end{lemma}

\begin{lemma} \label{lem-cross}
	Let $S$ be a set of points in $\mathbb{R}^2$ and $\mathcal{X}$ be a collection of ranges in $\mathbb{R}^2$.
	Suppose $(a,b),(a',b') \in \varPhi(S,\mathcal{X})$ are two pairs such that the segments $[a,b]$ and $[a',b']$ cross.
	Then there exists $X \in \mathcal{X}$ such that either $X \cap \{a,b,a',b'\} = \{a,b\}$ or $X \cap \{a,b,a',b'\} = \{a',b'\}$.
\end{lemma}

\subsection{Overview of key ideas and techniques} \label{sec-technique}
When $\varGamma$ is a polygon (possibly with holes), we solve the problem as follows.
First, we use a grid-based approach to reduce the $\mathcal{L}_\varGamma$-RCP problem to the RCP problem with wedge/co-wedge translation queries and the range-reporting problem with $\varGamma$-translation queries.
The range-reporting problem can be easily solved by again reducing to the wedge/co-wedge case.
Therefore, it suffices to study the RCP problem with wedge/co-wedge translation queries.
%Let $W$ be a fixed wedge.
For both wedge and co-wedge translation queries, we solve the problem by using candidate pairs.
Specifically, we store the candidate pairs and search for the answer among them.
In this approach, the critical point is the number of the candidate pairs, which determines the performance of our data structures.
For both wedge and co-wedge, we prove \textit{linear} upper bounds on the number of the candidate pairs.
Although the bounds are the same, the wedge case and co-wedge case require very different proofs, both of which are quite technical and may be of independent geometric interest.
These upper bounds and the above-mentioned reduction are our main technical contributions for the polygonal case.

When $\varGamma$ is a general convex body with a smooth boundary, we solve the problem as follows.
First, exploiting the smoothness of $\partial \varGamma$, we show that ``short'' candidate pairs (i.e., of length upper bounded by some constant $\tau$) cannot ``cross'' each other\footnote{We say two pairs $(a,b)$ and $(a',b')$ cross if the segments $[a,b]$ and $[a',b']$ cross.}. 
It immediately follows that there are only a linear number of short candidate pairs (because they form a planar graph).
We try to store these short candidate pairs in a data structure $\mathcal{D}_1$ such that the shortest one contained in any query $\varGamma_q$ can be found efficiently.
However, this is a nontrivial task, as $\varGamma$ is quite general here.
To this end, we reduce the task of ``searching for the shortest pair in $\varGamma_q$'' to several point-location queries for $q$ in planar subdivisions.
We bound the complexity of these subdivisions (and thus the cost of $\mathcal{D}_1$) by making geometric observations for convex translates and using properties of the pseudo-discs.
Using $\mathcal{D}_1$, we can answer any query $\varGamma_q$ in which the closest-pair is short.
What if the closest-pair in $\varGamma_q$ is long (i.e., of length greater than $\tau$)?
In this case, $\varGamma_q$ contains only $O(1)$ points by Lemma~\ref{lem-pigeon}.
Therefore, if $\mathcal{D}_1$ fails to find the answer, we can simply report the $O(1)$ points contained in $\varGamma_q$ and find the closest-pair by brute-force.
The range-reporting is done by point location in the $\leq k$-level of a pseudo-disc arrangement.
These are our main contributions for this part.

\section{Translation RCP queries for polygons} \label{sec-polygon}
Let $\varGamma$ be a fixed polygon (possibly with holes).
Assume the boundary of $\varGamma$ has no self-intersection\footnote{That is, the outer boundary and the boundaries of holes are disjoint simple cycles.}.
We investigate the $\mathcal{L}_\varGamma$-RCP problem (where the closest-pair is in terms of the Euclidean metric).
Throughout this section, $O(\cdot)$ hides constants depending on $\varGamma$.
Our main result is the following theorem, to prove which is the goal of this section.
\begin{theorem} \label{thm-plgres}
	Let $\varGamma$ be a fixed polygon \textnormal{(}possibly with holes\textnormal{)} in $\mathbb{R}^2$.
	Then there is an $O(n)$-space $\mathcal{L}_\varGamma$-RCP data structure with $O(\log n)$ query time.
\end{theorem}
Let $S$ be the given dataset in $\mathbb{R}^2$ of size $n$.
Suppose for convenience that the pairwise distances of the points in $S$ are distinct (so  the closest-pair in any subset of $S$ is unique).

\subsection{Reduction to (co-)wedge translation queries}
Our first step is to reduce a $\varGamma$-translation RCP query to several wedge/co-wedge translation RCP queries and a range-reporting query.
For a vertex $v$ of $\varGamma$ (either on the outer boundary or on the boundary of a hole), we define a wedge (or co-wedge) $W^v$ as follows.
Consider the two edges adjacent to $v$ in $\varGamma$.
These two edges define two (explementary) angles at $v$, one of which (say $\sigma$) corresponds to the interior of $\varGamma$ (while the other corresponds to the exterior of $\varGamma$).
Let $W^v$ be the (co-)wedge defined by $\sigma$ depending on whether $\sigma < \pi$ or $\sigma > \pi$.

%denote by $W^v$ the wedge with vertex $v$ and two bounding rays parallel to the two edges adjacent to $v$.
Let $\mathcal{W}_\varGamma = \{W^v: v \text{ is a vertex of } \varGamma\}$.
Without loss of generality, suppose that the outer boundary of $\varGamma$ consists of at least four edges, and so does the boundary of each hole\footnote{If this is not the case, we can add a new vertex at the midpoint of each edge to ``break'' it into two.}; with this assumption, no three edges of $\varGamma$ are pairwise adjacent.
For two edges $e$ and $e'$ of $\varGamma$, let $\text{dist}(e,e')$ denote the minimum distance between one point on $e$ and one point on $e'$.
Define $\delta = \min\{\text{dist}(e,e'): e \text{ and } e' \text{ are non-adjacent } \text{edges of } \varGamma\}$.
Clearly, $\delta$ is a positive constant depending on $\varGamma$ only.
Let $\Box$ be a square of side-length less than $\delta/\sqrt{2}$.
Due to the choice of $\delta$, for any $q \in \mathbb{R}^2$, $\Box$ cannot intersect two non-adjacent edges of $\varGamma_q$.
It follows that $\Box$ intersects at most two edges of $\varGamma_q$ (as no three edges of $\varGamma$ are pairwise adjacent); moreover, if $\Box$ intersects two edges, they must be adjacent.
Thus, $\Box \cap \varGamma_q = \Box \cap W_q$ for some $W \in \mathcal{W}_\varGamma$.

For a decomposable range-search problem (e.g. range reporting) on $S$, the above simple observation already allows us to reduce a $\varGamma$-translation query to (co-)wedge translation queries (roughly) as follows.
Let $G$ be a grid of width $\delta/2$ on the plane.
For a cell $\Box$ of $G$, we define $S_\Box = S \cap \Box$.
Due to the decomposability of the problem, to answer a query $\varGamma_q$ on $S$, it suffices to answer the query $\varGamma_q$ on $S_\Box$ for all $\Box$ that intersect $\varGamma_q$.
Since each cell $\Box$ of $G$ is a square of side-length $\delta/2$ (which is smaller than $\delta/\sqrt{2}$), we have $\Box \cap \varGamma_q = \Box \cap W_q$ for some $W \in \mathcal{W}_\varGamma$ and thus $S_\Box \cap \varGamma_q = S_\Box \cap W_q$.
In other words, the query $\varGamma_q$ on each $S_\Box$ is equivalent to a (co-)wedge translation query for some (co-)wedge $W \in \mathcal{W}_\varGamma$.
%We use $\varGamma_q \cap G$ to denote the set of the grid cells intersecting $\varGamma_q$; note that $|\varGamma_q \cap G| = O(1)$.
%For any $\varGamma_q \in \mathcal{L}_\varGamma$ and any cell $\Box$ of $G$, we have $\Box \cap \varGamma_q = \Box \cap W_q$ for some $W \in \mathcal{W}_\varGamma$, which implies $S_\Box \cap \varGamma_q = S_\Box \cap W_q$.
%Since $S_\Box \cap \varGamma_q = S_\Box \cap W_q$ for some 
Applying this idea to range-reporting, we conclude the following.
%In this way, we reduce a polygonal translation query to several (co-)wedge translation queries.
%For instance, suppose we want to do range reporting on $S$ for the range space $\mathcal{L}$.
%Let $S \cap G$ denote the set of the nonempty cells of $G$; note that $|S \cap G| = O(n)$.
%For each $\Box \in S \cap G$ and each $W \in \mathcal{W}_\varGamma$, we build a range-reporting data structure on $S_\Box$ for the range space $\{W_q: q \in \mathbb{R}^2\}$; there exists such a data structure with $O(|S_\Box|)$ space and $O(\log |S_\Box| + k)$ query time, where $k$ is the number of the reported points.
\begin{lemma} \label{lem-rangerep}
	There exists an $O(n)$-space range-reporting data structure for $\varGamma$-translation queries, which has an $O(\log n +k)$ query time, where $k$ is the number of the reported points.
\end{lemma}
However, the above argument fails for a non-decomposable range-search problem, since when the problem is non-decomposable, we are not able to recover efficiently the global answer even if the answer in each cell is known.
Unfortunately, our RCP problem belongs to this category.
Therefore, more work is required to do the reduction.
We shall take advantage of our observation in Lemma~\ref{lem-pigeon}.
We still lay a planar grid $G$.
But this time, we set the width of $G$ to be $\delta/4$.
A \textit{quad-cell} $\boxplus$ of $G$ is a square consisting of $2 \times 2$ adjacent cells of $G$.
For a quad-cell $\boxplus$ of $G$, let $S_\boxplus = S \cap \boxplus$.
Note that the side-length of a quad-cell of $G$ is $\delta/2$, and each cell of $G$ is contained in exactly four quad-cells of $G$, so is each point in $S$.
Consider a query range $\varGamma_q \in \mathcal{L}_\varGamma$.
The following observation follows from Lemma~\ref{lem-pigeon}.
\begin{lemma} \label{lem-qcell}
	For a a quad-cell $\boxplus$ of $G$ such that $|S_\boxplus \cap \varGamma_q| \geq 2$, let $\phi_\boxplus$ be the closest-pair in $S_\boxplus \cap \varGamma_q$.
	Define $\phi^*$ as the shortest element among all $\phi_\boxplus$.
	If the length of $\phi^*$ is at most $\delta/4$, then $\phi^*$ is the closest-pair in $S \cap \varGamma_q$; otherwise $|S \cap \varGamma_q| = O(1)$.
\end{lemma}

\noindent
Using the above observation, we are able to do the reduction.
%First, we compute $\phi^*$ in the above lemma.
%To this end, we need to compute $\phi_\boxplus$ for all quad-cells $\boxplus$ of $G$ that intersect $\varGamma_q$.
%Note that $\boxplus \cap \varGamma_q = \boxplus \cap W_q$ for some $W \in \mathcal{W}_\varGamma$, as the side-length of $\boxplus$ is $\delta/2$.
%Therefore, $\phi_\boxplus$ can be computed by answering a (co-)wedge translation RCP query on $S_\boxplus$ for some (co-)wedge $W \in \mathcal{W}_\varGamma$.
%If the length of $\phi^*$ is at most $\delta/4$, then we are done, as $\phi^*$ is just the closest-pair in $S \cap \varGamma_q$ by Lemma~\ref{lem-qcell}.
%Otherwise, Lemma~\ref{lem-qcell} implies $|S \cap \varGamma_q| = O(1)$.
%So we can report all the points in $S \cap \varGamma_q$ and compute the closest-pair by brute-force.
%Along with this idea, the query $\varGamma_q$ is reduced to several (co-)wedge translation RCP queries for (co-)wedges in $\mathcal{W}_\varGamma$ and a $\varGamma$-translation range-reporting query.
%We have already seen in Lemma~\ref{lem-rangerep} that range-reporting for $\varGamma$-translation queries can be solved optimally, thus the problem is actually reduced to handling wedge and co-wedge translation RCP queries.
%Formally, we conclude the following.
\begin{theorem} \label{thm-reduction}
	%Let $W$ be a \textnormal{(}co-\textnormal{)}wedge in $\mathbb{R}^2$, and $\mathcal{L}_W = \{W_q: q \in \mathbb{R}^2\}$.
	Let $f,g: \mathbb{N} \rightarrow \mathbb{N}$ be increasing functions where $f(a+b) \geq f(a)+f(b)$.
	If for any $W \in \mathcal{W}_\varGamma$ there is an $O(f(n))$-space $\mathcal{L}_{W}$-RCP data structure with $O(g(n))$ query time, then there is an $O(f(n)+n)$-space $\mathcal{L}_\varGamma$-RCP data structure with $O(g(n)+\log n)$ query time.
\end{theorem}
\textit{Proof.}
For a quad-cell $\boxplus$ of $G$, let $m_\boxplus$ be the number of the points in $S_\boxplus$.
First, we notice that there are $O(n)$ quad-cells $\boxplus$ of $G$ such that $m_\boxplus>0$ since each point in $S$ is contained in at most four quad-cells; we call them \textit{nonempty} quad-cells.
For each nonempty quad-cell $\boxplus$ and each $W \in \mathcal{W}_\varGamma$, we build an $\mathcal{L}_{W}$-RCP data structure on $S_\boxplus$; by assumption, this data structure uses $O(f(m_\boxplus))$ space.
Now observe that $m_\boxplus \leq n$ for all $\boxplus$ and $\sum m_\boxplus \leq 4n$.
From the condition $f(a+b) \geq f(a)+f(b)$, it follows that $\sum f(m_\boxplus) = O(f(n))$.
Since $|\mathcal{W}_\varGamma| = O(1)$, the total space cost of these data structures is $O(f(n))$.
Besides these data structures, we also build a range-reporting data structure on $S$ for $\varGamma$-translation queries.
As argued in Lemma~\ref{lem-rangerep}, this data structure uses $O(n)$ space.
%Therefore, the overall space complexity is $O(f(n)+n)$.

To answer a query $\varGamma_q \in \mathcal{L}_\varGamma$, we first find all nonempty quad-cells of $G$ that intersect $\varGamma_q$.
The number of these quad-cells is $O(1)$, as it is bounded by $O(\Delta^2 / \delta^2)$ where $\Delta$ is the diameter of $\varGamma$.
These quad-cells can be found in $O(\log n)$ time (see \cite{xue2018searching}).
For each such quad-cell $\boxplus$, we find $W \in \mathcal{W}_\varGamma$ such that $\boxplus \cap \varGamma_q = \boxplus \cap W_q$ and query the $\mathcal{L}_W$-RCP data structure built on $S_\boxplus$ to obtain the closest-pair $\phi_\boxplus$ in $S_\boxplus \cap \varGamma_q$, which takes $O(g(m_\boxplus))$ time.
Since only $O(1)$ quad-cells are considered, the time for this step is $O(g(n))$.
Once these $\phi_\boxplus$ are computed, we take the shortest element $\phi^*$ among them.
%Until now, it takes $O(g(n))$ time.
If the length of $\phi^*$ is at most $\delta/4$, then $\phi^*$ is the closest-pair in $S \cap \varGamma_q$ by Lemma~\ref{lem-qcell} and we just report $\phi^*$.
Otherwise, $|S \cap \varGamma_q| = O(1)$ by Lemma~\ref{lem-qcell}.
We then compute the $O(1)$ points in $S \cap \varGamma_q$ using the range-reporting data structure, and compute the closest-pair in $S \cap \varGamma_q$ by brute-force (in constant time).
Since the query time of the range-reporting data structure is $O(\log n+k)$ and $k=O(1)$ here, the overall query time is $O(g(n)+\log n)$, as desired.
\hfill $\Box$
\medskip

\noindent
By the above theorem, it now suffices to give efficient RCP data structures for wedge and co-wedge translation RCP queries.
We resolve these problems in the following two sections.

\subsection{Handling wedge translation queries}
\label{Handling wedge translation queries}
%According to the previous section, it now suffices to establish an efficient solution for wedge translation RCP queries.
Let $W$ be a fixed wedge in $\mathbb{R}^2$ and $\theta \in (0,\pi)$ be the angle of $W$.
We denote by $r$ and $r'$ the two branches of $W$.
For convenience, assume the vertex of $W$ is the origin, and thus the vertex of a $W$-translate $W_p$ is the point $p$.
%$r = \{(t,0): t \geq 0\}$ and $r' = \{(\alpha t,t): t \geq 0\}$ for some $\alpha \in \mathbb{R}$.
In this section, we shall give an $O(n)$-space $\mathcal{L}_W$-RCP data structure with $O(\log n)$ query time.

The key ingredient of our result is a nontrivial linear upper bound for the number of the candidate pairs in $S$ with respect to $\mathcal{L}_W$.
This generalizes a result in \cite{gupta2014data}, and requires a much more technical proof.
Before working on the proof, we first establish an easy fact.
\begin{lemma} \label{lem-smallest}
	Let $A \subseteq \mathbb{R}^2$ be a finite set.
	There exists a \textnormal{(}unique\textnormal{)} smallest $W$-translate \textnormal{(}under the $\subseteq$-order\textnormal{)} that contains $A$.
	Furthermore, a $W$-translate is the smallest $W$-translate containing $A$ iff it contains $A$ and its two branches both intersect $A$.
\end{lemma}

\begin{figure}[h]
	\centering
	\begin{subfigure}[b]{4.5cm}
	    \includegraphics[height=1.1cm]{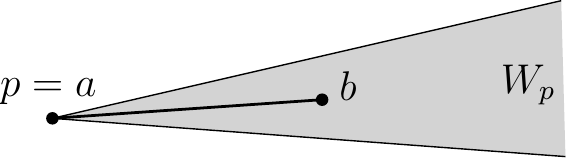}
	    \caption{Flat (as $p=a$)}
	\end{subfigure}
	%\hspace{0.3cm}
	\begin{subfigure}[b]{4.3cm}
	    \includegraphics[height=1.3cm]{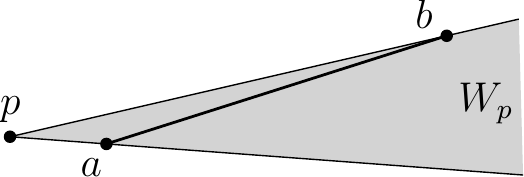}
	    \caption{Flat (as $\angle apb > \angle pba$)}
	\end{subfigure}	
	%\hspace{0.3cm}
	\begin{subfigure}[b]{4.5cm}
	    \includegraphics[height=1.3cm]{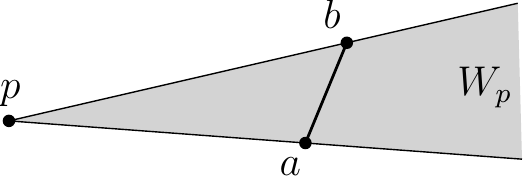}
	    \caption{Steep}
	\end{subfigure}
	\caption{Examples of flat and steep pairs (the wedge $W_p$ shown in the figures is the smallest $W$-translate containing $\{a,b\}$ described in Lemma~\ref{lem-smallest}).}
	\label{fig-flat&steep}
\end{figure}

\noindent
We notice that if $\phi = (a,b)$ is a pair of points in $S$ and $W_p$ is the smallest $W$-translate containing $\{a,b\}$ described in Lemma~\ref{lem-smallest}, then $\phi \in \varPhi(S,\mathcal{L}_W)$ iff $\phi$ is the closest-pair in $S \cap W_p$.
Using Lemma~\ref{lem-smallest}, we define the following notions.
\begin{definition}
Let $\phi = (a,b)$ be a pair of points in $\mathbb{R}^2$, and $W_p$ be the smallest $W$-translate containing $\{a,b\}$ described in Lemma~\ref{lem-smallest}.
If $p \notin \{a,b\}$ and the smallest angle of the triangle $\triangle p a b$ is $\angle a p b$, then we say $\phi$ is \textbf{steep}; otherwise, we say $\phi$ is \textbf{flat}.
See Figure~\ref{fig-flat&steep} for examples.	
\end{definition}
Our first observation is the following.
\begin{lemma} \label{lem-flatnoncrossing}
	If two candidate pairs $\phi,\phi' \in \varPhi(S,\mathcal{L}_W)$ cross, then either $\phi$ or $\phi'$ is steep.
\end{lemma}
\textit{Proof.}
Suppose $\phi = (a,b)$ and $\phi' = (a',b')$.
Since $\phi$ and $\phi'$ cross, by Lemma~\ref{lem-cross} there exists some $W_t \in \mathcal{L}_W$ whose intersection with $\{a,b,a',b'\}$ is either $\{a,b\}$ or $\{a',b'\}$; assume $W_t \cap \{a,b,a',b'\} = \{a,b\}$.
Let $p \in \mathbb{R}^2$ be the point such that $W_p$ is the smallest $W$-translate containing $\{a,b\}$.
It follows that $W_p \cap \{a,b,a',b'\} = \{a,b\}$, because $W_p \subseteq W_t$.
Let $c$ be the intersection point of the segments $[a,b]$ and $[a',b']$.
Since $a,b \in W_p$ and $W_p$ is convex, $c \in W_p$.
The two endpoints $a',b'$ of the segment $[a',b']$ are not contained in $W_p$ (by assumption), but $c \in W_p$.
\begin{figure}[h]
	\centering
	\begin{subfigure}[b]{5cm}
	    \includegraphics[height=1.8cm]{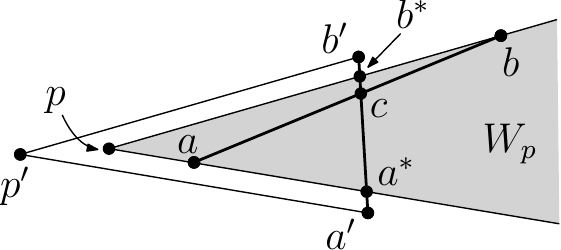}
	    \caption{The points $a^*$, $b^*$, and $p'$}
	    \label{fig-noncr1}
	\end{subfigure}	
	\hspace{0.5cm}
	\begin{subfigure}[b]{5cm}
	    \includegraphics[height=1.8cm]{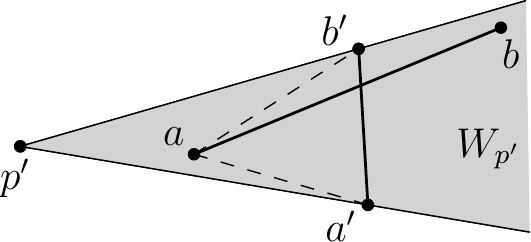}
	    \caption{Illustrating why $(a',b')$ is steep}
	    \label{fig-noncr2}
	\end{subfigure}
	\caption{Illustrating Lemma~\ref{lem-flatnoncrossing}.}
	\label{fig-noncr}
\end{figure}
Hence, the segment $[a',b']$ intersects the boundary of $W_p$ at two points, say $a^*$ and $b^*$; assume $a^*$ (resp., $b^*$) is the point adjacent to $a'$ (resp., $b'$).
Clearly, there exists a unique point $p' \in \mathbb{R}^2$ such that $\triangle p a^* b^* \subseteq \triangle p' a' b'$ and $\triangle p a^* b^*$ is similar to $\triangle p' a' b'$.
See Figure~\ref{fig-noncr1}.
It is easy to see that $W_{p'}$ is the smallest $W$-translate containing $\{a',b'\}$.
Indeed, $W_{p'}$ just corresponds to the angle $\angle a' p' b'$, so $a'$ and $b'$ lie on the two branches of $W_{p'}$ respectively (see Figure~\ref{fig-noncr2}).
Thus, by the criterion given in Lemma~\ref{lem-smallest}, $W_{p'}$ is the smallest $W$-translate containing $\{a',b'\}$.
Now we have $p \in W_{p'}$, which implies $W_p \subseteq W_{p'}$ and $a,b,a',b' \in W_{p'}$.
Since the segments $[a,b]$ and $[a',b']$ cross, one of $a$ and $b$ must lie in the triangle $\triangle p' a' b'$, say $a \in \triangle p' a' b'$.
Note that $\phi' = (a',b')$ is the closest-pair in $W_{p'}$, thus $\text{dist}(a',b')<\text{dist}(a,a')$ and $\text{dist}(a',b')<\text{dist}(a,b')$.
It follows that $\angle a' a b' < \angle a b' a'$ and $\angle a' a b' < \angle a a' b'$.
We further observe that $\angle a a' b' < \angle p' a' b'$ and $\angle a b' a' < \angle p' b' a'$, and hence $\angle a' a b' > \angle a' p' b'$.
Thus, we have $\angle a' p' b' < \angle p' a' b'$ and $\angle a' p b' < \angle p' b' a'$, i.e., $\angle a' p' b'$ is the smallest angle of the triangle $\triangle p' a' b'$.
As a result, $\phi'$ is steep.
\hfill $\Box$
\smallskip

\noindent
Lemma~\ref{lem-flatnoncrossing} implies that the flat candidate pairs in $\varPhi(S,\mathcal{L}_W)$ do not cross each other.
Therefore, the segments corresponding to the flat candidate pairs are edges of a planar graph with vertices in $S$, which gives a linear upper bound for the number of flat candidate pairs.

It now suffices to bound the number of steep candidate pairs in $\varPhi(S,\mathcal{L}_W)$.
Unfortunately, two steep candidate pairs (or even one steep candidate pair and one flat candidate pair) can cross, making the above non-crossing argument fail.
Therefore, we need some new ideas.
\begin{definition}
    Two pairs $\phi,\phi' \in \varPhi(S,\mathcal{L}_W)$ are \textbf{adjacent} if we can write $\phi = (a,b)$ and $\phi' = (a,b')$ such that $b \neq b'$; we call $\angle b a b'$ the \textbf{angle} between $\phi$ and $\phi'$, denoted by $\textnormal{ang}(\phi,\phi')$.
\end{definition}
\begin{lemma} \label{lem-steepbigangle}
	For adjacent $\phi,\phi' \in \varPhi(S,\mathcal{L}_W)$, if $\phi$ and $\phi'$ are both steep, then $\textnormal{ang}(\phi,\phi') \geq \theta$.
\end{lemma}

\begin{figure}[b]
	\centering
	\begin{subfigure}[b]{4.5cm}
	    \includegraphics[height=1.8cm]{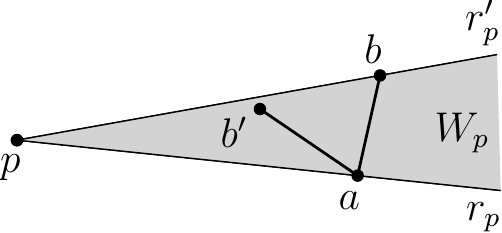}
	    \caption{Case~1 when $b' \in \triangle pab$}
	    \label{fig-bigang1}
	\end{subfigure}	
	%\hspace{0.5cm}
	\begin{subfigure}[b]{4.5cm}
	    \includegraphics[height=1.8cm]{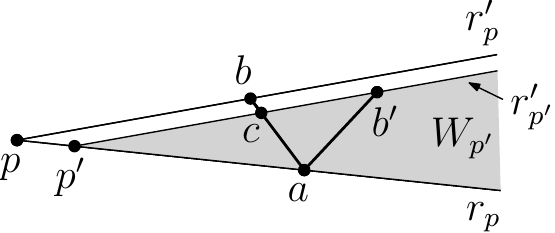}
	    \caption{Case~1 when $b' \notin \triangle pab$}
	    \label{fig-bigang2}
	\end{subfigure}		
	%\hspace{0.5cm}
	\begin{subfigure}[b]{4.5cm}
	    \includegraphics[height=1.8cm]{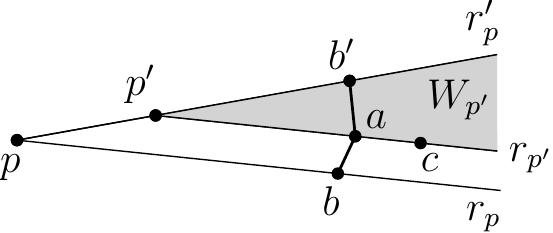}
	    \caption{Case~2}
	    \label{fig-bigang3}
	\end{subfigure}		
	\caption{An illustration of the proof of Lemma~\ref{lem-steepbigangle}.}
	\label{fig-bigang}
\end{figure}
\noindent
\textit{Proof.}
Since $\phi$ and $\phi'$ are adjacent, we can write $\phi = (a,b)$ and $\phi' = (a,b')$.
%Assume $\phi$ and $\phi'$ are both steep.
Let $p \in \mathbb{R}^2$ be the point such that $W_p$ is the smallest $W$-translate containing $\{a,b,b'\}$ described in Lemma~\ref{lem-smallest}.
We first notice that $p \notin \{a,b,b'\}$.
Indeed, if $p \in \{a,b\}$, then $W_p$ is the smallest $W$-translate containing $\{a,b\}$ by the criterion given in Lemma~\ref{lem-smallest}.
However, $\phi = (a,b)$ is steep by assumption, hence $p \notin \{a,b\}$, which results in a contradiction.
So we have $p \notin \{a,b\}$.
For the same reason, we have $p \notin \{a,b'\}$ and thus $p \notin \{a,b,b'\}$.
We denote by $r_p$ and $r'_p$ the $r$-branch and $r'$-branch of $W_p$, respectively.
To prove the lemma, we consider two cases separately: $a$ is on the boundary of $W_p$ or $a$ is in the interior of $W_p$.
\smallskip

\noindent
\textbf{[Case 1]}
Assume $a$ is on the boundary of $W_p$.
Since $p \notin \{a,b,b'\}$, we have $a \neq p$.
Thus $a$ must lie on exactly one of $r_p$ and $r'_p$, say $a \in r_p$.
Because $W_p$ is the smallest $W$-translate containing $\{a,b,b'\}$, one of $b$ and $b'$ must lie on $r'_p$, by the criterion given in Lemma~\ref{lem-smallest}.
Without loss of generality, assume $b \in r'_p$.
Using the criterion in Lemma~\ref{lem-smallest} again, we see that $W_p$ is also the smallest $W$-translate containing $\{a,b\}$.
Thus, $\phi$ is the closest-pair in $S \cap W_p$ and in particular we have $\text{dist}(a,b)<\text{dist}(b,b')$.
It follows that $\angle ab'b < \angle bab' = \textnormal{ang}(\phi,\phi')$.
If $b' \in \triangle pab$, we are done, because in this case we have $\angle ab'b \geq \angle apb = \theta$ and thus $\textnormal{ang}(\phi,\phi') = \angle bab' > \angle ab'b \geq \angle apb = \theta$.
See Figure~\ref{fig-bigang1} for an illustration of this case.
Next, assume $b' \notin \triangle pab$.
This case is presented in Figure~\ref{fig-bigang2}.
Let $p' \in \mathbb{R}^2$ be the point such that $W_{p'}$ is the smallest $W$-translate containing $\{a,b'\}$; thus, we have $W_{p'} \subseteq W_p$ and in particular $p' \in W_p$.
Furthermore, $p'$ must lie on the segment $[p,a]$, as $a \in W_{p'}$.
%We denote by $r_2'$ the bounding ray of $W_{p'}$ that is parallel to $r_2$, and by $c$ the intersection point of $r_2'$ and the segment $[a,b]$.
Since $\phi' = (a,b')$ is steep, $a$ and $b'$ lie on the two branches of $W_{p'}$ respectively, and $\angle p'b'a > \angle ap'b' = \theta$.
Clearly, $b'$ lies on the $r'$-branch of $W_{p'}$, which we denote by $r'_{p'}$.
Let $c$ be the intersection point of $r'_{p'}$ and the segment $[a,b]$.
See Figure~\ref{fig-bigang2}.
%Because $\phi' = (a,b')$ is steep, $\angle p'b'a > \angle ap'b' = \theta$.
We then have
\begin{equation*}
\angle ab'b = \angle ab'c + \angle cb'b \geq \angle ab'c = \angle pb'a > \angle ap'b' = \theta.
\end{equation*}
Using the fact $\angle ab'b < \angle bab' = \textnormal{ang}(\phi,\phi')$ obtained before, we conclude $\textnormal{ang}(\phi,\phi') \geq \theta$.
%See Figure~\ref{fig-bigang2} for an illustration of this case.
%Consider the line $l$ through $a$ and $b'$.
%Clearly, $l$ intersects $r'_p$ at a point $b^*$ such that $\triangle p' a b' \sim \triangle p a b^*$.
\smallskip

\noindent
\textbf{[Case 2]}
Assume $a$ is in the interior of $W_p$.
This case is presented in Figure~\ref{fig-bigang3}.
%This case is actually easier than the previous case.
By the criterion given in Lemma~\ref{lem-smallest}, both $r_p$ and $r'_p$ intersect $\{a,b,b'\}$.
Since $p \notin \{a,b,b'\}$ and $a \notin \partial W_p$, $b$ and $b'$ must lie on the two branches of $W_p$ respectively, say $b \in r_p$ and $b' \in r'_p$.
Let $p' \in \mathbb{R}^2$ be the point such that $W_{p'}$ is the smallest $W$-translate containing $\{a,b'\}$; thus, we have $W_{p'} \subseteq W_p$ and in particular $p' \in W_p$.
Furthermore, $p'$ must lie on the segment $[p,b']$, as $b' \in W_{p'}$.
Since $\phi' = (a,b')$ is steep, $a$ and $b'$ are on the two branches of $W_{p'}$ respectively, and $\angle p'ab' > \angle ap'b' = \theta$.
Clearly, $a$ lies on the $r$-branch of $W_{p'}$, which we denote by $r_{p'}$.
Take a point $c \in r_{p'}$ such that $a$ lies on the segment $[p',c]$.
See Figure~\ref{fig-bigang3}.
Consider the two supplementary angles $\angle p'ab'$ and $\angle b'ac$.
%We claim that both of them are greater than $\theta$.
We have $\angle p'ab' > \theta$ as argued before.
Also, we have $\angle b'ac > \theta$, because $\angle b'ac = \angle ap'b' + \angle p'b'a$ and $\angle ap'b' = \theta$.
Note that the segments $[a,b']$ and $[a,b]$ are on opposite sides of $r_{p'}$.
Therefore, either $\angle bab' = \angle p'ab' + \angle p'ab$ or $\angle bab' = \angle b'ac + \angle bac$.
In either of the two cases, we have $\textnormal{ang}(\phi,\phi') = \angle bab' > \theta$ because $\angle p'ab' > \theta$ and $\angle b'ac > \theta$.
%See Figure~\ref{fig-bigang3} for an illustration.
\hfill $\Box$
\smallskip

\noindent
For a point $a \in S$, consider the subset $\varPsi_a \subseteq \varPhi(S,\mathcal{L}_W)$ consisting of all steep candidate pairs having $a$ as one point.
We claim $|\varPsi_a| = O(1)$.
Suppose $\varPsi_a = \{\psi_1,\dots,\psi_r\}$ where $\psi_i = (a,b_i)$ and $b_1,\dots,b_r$ are sorted in polar-angle order around $a$.
By Lemma~\ref{lem-steepbigangle}, $\text{ang}(\psi_i,\psi_j) \geq \theta$ for any distinct $i,j \in \{1,\dots,r\}$.
Since $\sum_{i=1}^{r-1} \text{ang}(\psi_i,\psi_{i+1}) \leq 2 \pi$, we have $r \leq 2\pi/\theta+1 = O(1)$.
As such, $\sum_{a \in S} |\varPsi_a| = O(n)$, implying that the number of steep candidate pairs is linear.
As the numbers of flat and steep candidate pairs are both linear, we conclude the following.
\begin{lemma} \label{lem-linearwedcand}
	$|\varPhi(S,\mathcal{L}_W)| = O(n)$, where $n = |S|$.
\end{lemma}

\noindent
Suppose $\varPhi(S,\mathcal{L}_W) = \{\phi_1,\dots,\phi_m\}$ where $\phi_i = (a_i,b_i)$ and $\phi_1,\dots,\phi_m$ are sorted in increasing order of their lengths.
We have $m = O(n)$ by Lemma~\ref{lem-linearwedcand}.
Now we only need a data structure which can report, for a query $W_q \in \mathcal{L}_W$, the smallest $i$ such that $a_i,b_i \in W_q$ (note that $\phi_i$ is the closest-pair in $S \cap W_q$).
We design this data structure as follows.
Let $\tilde{W} = \{(x,y):(-x,-y) \in W\}$, which is a wedge obtained by rotating $W$ around the origin with angle $\pi$.
For a point $p \in \mathbb{R}^2$, it is clear that $a_i,b_i \in W_p$ iff $p \in \tilde{W}_{a_i} \cap \tilde{W}_{b_i}$.
Since the intersection of finitely many $\tilde{W}$-translates is a $\tilde{W}$-translate, we may write $\tilde{W}_{a_i} \cap \tilde{W}_{b_i} = \tilde{W}_{c_i}$ for some $c_i \in \mathbb{R}^2$.
It follows that $\phi_i$ is contained in $W_p$ iff $p \in \tilde{W}_{c_i}$.
By successively overlaying $\tilde{W}_{c_1},\dots,\tilde{W}_{c_m}$, we obtain a planar subdivision whose cells are $\varSigma_1,\dots,\varSigma_m$ where $\varSigma_i = \tilde{W}_{c_i} \backslash \bigcup_{j=1}^{i-1} \tilde{W}_{c_j}$.
This subdivision has $O(m)$ complexity as overlaying a new $\tilde{W}$-translate can create at most two new vertices.
The answer for a query $W_q$ is $i$ iff $q \in \varSigma_i$.
Therefore, the problem can be solved by building on the subdivision an $O(m)$-space point-location data structure with $O(\log m)$ query time.
Since $m = O(n)$, we have the following conclusion.

\begin{theorem} \label{thm-wedds}
	There is an $O(n)$-space $\mathcal{L}_W$-RCP data structure with $O(\log n)$ query time.
\end{theorem}
%Combining Theorem~\ref{thm-reduction} and~\ref{thm-wedds}, Theorem~\ref{thm-plgres} is then proved.

\subsection{Handling co-wedge translation queries}
Let $C$ be a fixed co-wedge in $\mathbb{R}^2$ and $W$ be the \textit{complementary} wedge of $C$, i.e., the closure of $\mathbb{R}^2 \backslash C$.
We denote by $r$ and $r'$ the two branches of $C$ (and also of $W$).
For convenience, assume $r = \{(t,0): t \geq 0\}$ and $r' = \{(\alpha t,t): t \geq 0\}$ for some $\alpha \in \mathbb{R}$.
With this assumption, the vertex of $C$ (resp., $W$) is the origin and the vertex of $C_p$ (resp., $W_p$) is $p$ for all $p \in \mathbb{R}^2$.
%Let $H$ be the downward-open halfplane bounded by $r$ and $H'$ be the leftward-open halfplane bounded by $r'$.
%Then $C = H \cup H'$ and $C_p = H_p \cup H_p'$ for all $p \in \mathbb{R}^2$.
In this section, we present an $O(n)$-space $\mathcal{L}_C$-RCP data structure with $O(\log n)$ query time.

Similar to the wedge case, the key step here is to establish a linear upper bound for $|\varPhi(S,\mathcal{L}_C)|$.
However, the techniques used here are very different.
First of all, we exclude from $\varPhi(S,\mathcal{L}_C)$ the candidate pairs with respect to halfplanes.
Let $\mathcal{H}$ be the collection of halfplanes, and $\varPhi^* = \varPhi(S,\mathcal{L}_C) \backslash \varPhi(S,\mathcal{H})$.
It was shown in \cite{abam2009power} that $|\varPhi(S,\mathcal{H})| = O(n)$.
Therefore, it suffices to prove that $|\varPhi^*| = O(n)$.
For a pair $\phi=(a,b) \in \varPhi^*$, define its \textit{associated} $C$-translate, $\text{Ass}(\phi)$, as the smallest $W$-translate containing $\{a,b\}$ (Lemma~\ref{lem-smallest}).
%if $W_p \in \mathcal{L}_W$ is , we call $C_p$ the \textit{associated} $C$-translate of $\phi$.
The pairs in $\varPhi^*$ and their associated $C$-translates has the following property.
\begin{lemma} \label{lem-associate}
    Let $\phi=(a,b) \in \varPhi^*$ and $C_p = \textnormal{Ass}(\phi) \in \mathcal{L}_C$.
    Then $p \notin \{a,b\}$ and $a,b$ lie on the two branches of $C_p$ respectively.
    Furthermore, $\phi$ is the closest-pair in $S \cap C_p$.
\end{lemma}

\noindent
Consider a pair $\phi \in \varPhi^*$ and its associated $C$-translate $C_p = \text{Ass}(\phi)$.
By Lemma~\ref{lem-associate}, one point of $\phi$ lies on the $r$-branch of $C_p$ and the other lies on the $r'$-branch of $C_p$; we call them the $r$-\textit{point} and $r'$-\textit{point} of $\phi$, respectively.
Let $R \subseteq S$ (resp., $R' \subseteq S$) be the subset consisting of all the $r$-points (resp., $r'$-points) of the pairs in $\varPhi^*$.
\begin{lemma} \label{lem-bipartite}
    We have $R \cap R' = \emptyset$, and thus the graph $G = (S,\varPhi^*)$ is bipartite.
\end{lemma}

\begin{figure}[h]
    \centering
    \includegraphics[height=3.5cm]{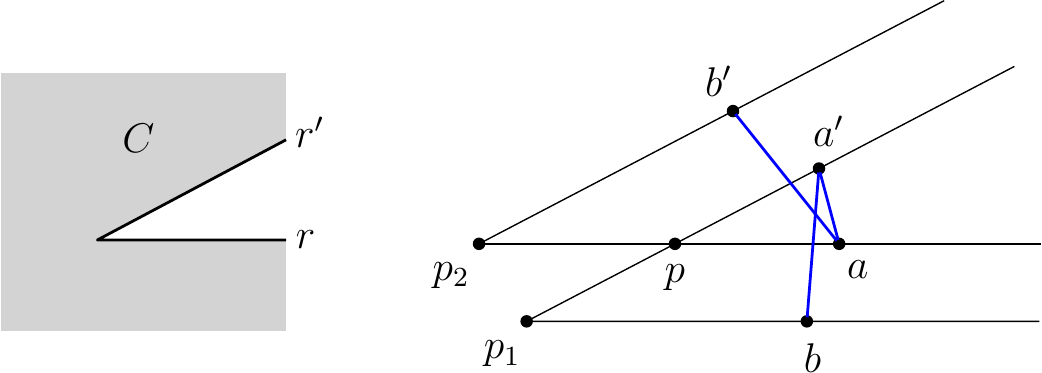}
    \caption{Illustration of the proof of Lemma~\ref{lem-acyclic}}
    \label{fig-prfacyc}
\end{figure}

\noindent
For a pair $\phi = (a,b) \in \varPhi^*$ where $a \in R$ and $b \in R'$, we define a vector $\mathbf{v}_\phi = \overrightarrow{ab}$.
Our key lemma is the following.
Let $\text{ang}(\cdot,\cdot)$ denote the angle between two vectors.
\begin{lemma} \label{lem-acyclic}
    Let $\varPsi \subseteq \varPhi^*$ be a subset such that $\textnormal{ang}(\mathbf{v}_\psi,\mathbf{v}_{\psi'}) \leq \pi/4$ for all $\psi,\psi' \in \varPsi$.
    Then the graph $G_\varPsi = (S,\varPsi)$ is acyclic, and in particular $|\varPsi| = O(n)$.
\end{lemma}
\textit{Proof.}
Suppose there is a cycle in $G_\varPsi$.
Let $\psi = (a,a')$ be the shortest edge in the cycle where $a \in R$ and $a' \in R'$.
Let $\psi_1 = (b,a') \in \varPsi$ and $\psi_2 = (a,b') \in \varPsi$ be the two adjacent edges of $\psi$ in the cycle (so $b \in R$ and $b' \in R'$).
Then $|\psi| < |\psi_1|$ and $|\psi| < |\psi_2|$.
Let $C_p,C_{p_1},C_{p_2}$ be the associated $C$-translates of $\psi,\psi_1,\psi_2$, respectively.
Since $r = \{(t,0): t \geq 0\}$ is a horizontal ray by our assumption, $b$ and $p_1$ have the same $y$-coordinate (as $b$ is on the $r$-branch of $C_{p_1}$).
Similarly, $a,p,p_2$ have the same $y$-coordinate.
See Figure~\ref{fig-prfacyc} for an illustration.
We first show that $C_{p_1} \subseteq C_p$ and $a,b,b' \in C_{p_2}$.
Note that $a'$ is on the $r'$-branches of both $C_p$ and $C_{p_1}$.
Thus, $a',p,p_1$ are collinear and the line through them is parallel to $r'$.
%Similarly, $a$ is on the $r'$-branch of $C_p$ and also the $r'$-branch of $C_{p_1}$.
This further implies that either $C_p \subseteq C_{p_1}$ or $C_{p_1} \subseteq C_p$.
Because $|\psi|<|\psi_1|$ and $\psi_1$ is the closest-pair in $C_{p_1}$ by Lemma~\ref{lem-associate}, $C_{p_1}$ does not contain $\psi$ and thus $C_{p_1} \subseteq C_p$ (as $C_p$ contains $\psi$ by Lemma~\ref{lem-associate}); in fact, $p$ is on the $r'$-branch of $C_{p_1}$.
It follows that $p_1$ is below $p$ (since $r' = \{(\alpha t,t): t \geq 0\}$ is an upward ray).
As argued before, $b$ has the same $y$-coordinate as $p_1$ and $p_2$ has the same $y$-coordinate as $p$.
Hence, $b$ is below $p_2$ and $b \in C_{p_2}$.
We have $a,b' \in C_{p_2}$ by Lemma~\ref{lem-associate}.
Now we see $a',b,b' \in C_{p_1}$.
Using the same argument symmetrically, we can prove $C_{p_2} \subseteq C_p$ and $a',b,b' \in C_{p_1}$.
Because $C_{p_1} \subseteq C_p$ and $C_{p_2} \subseteq C_p$, we have $a,a',b,b' \in C_{p_1} \cup C_{p_2} \subseteq C_p$.

Since $a,a',b,b' \in C_p$ and $\phi$ is the closest-pair in $C_p$ by Lemma~\ref{lem-associate}, we have $|\phi|<\text{dist}(a,b)$ and hence $\angle aba' < \angle aa'b = \text{ang}(\mathbf{v}_\psi,\mathbf{v}_{\psi_1}) \leq \pi/4$.
It follows that
\begin{equation*}
    \text{ang}(\overrightarrow{ba},\mathbf{v}_\psi) \leq \text{ang}(\overrightarrow{ba},\mathbf{v}_{\psi_1}) + \text{ang}(\mathbf{v}_\psi,\mathbf{v}_{\psi_1}) = \angle aba' + \text{ang}(\mathbf{v}_\psi,\mathbf{v}_{\psi_1}) < \pi/2.
\end{equation*}
From this we can further deduce that
\begin{equation*}
    \text{ang}(\overrightarrow{bb'},\mathbf{v}_\psi) = \text{ang}(\overrightarrow{ba}+\mathbf{v}_{\psi_2},\mathbf{v}_\psi)
    \leq \max\{\text{ang}(\overrightarrow{ba},\mathbf{v}_\psi),\text{ang}(\mathbf{v}_{\psi_2},\mathbf{v}_\psi)\} < \pi/2.
\end{equation*}
Next, we shall establish an inequality that contradicts the above inequality.
Let $l$ be the bisector of the segment $[b,b']$.
Since $a',b,b' \in C_{p_1}$ and $\psi_1$ is the closest-pair in $C_{p_1}$, we have $\text{dist}(a',b') > \text{dist}(b,a')$, i.e., $a'$ is on the same side of $l$ as $b$.
Using the same argument symmetrically, we can deduce that $a$ is on the same side of $l$ as $b'$.
Since $l$ is the bisector of $[b,b']$, this implies $\text{ang}(\overrightarrow{bb'},\mathbf{v}_\psi) = \text{ang}(\overrightarrow{bb'},\overrightarrow{aa'}) > \pi/2$, which is a contradiction.
Therefore, we see that $G_\varPsi$ is acyclic.
\hfill $\Box$
\smallskip

\noindent
With the above lemma in hand, it is quite straightforward to prove $|\varPhi^*| = O(n)$.
We evenly separate the plane into 8 sectors $K_1,\dots,K_8$ around the origin.
Define $\varPsi_i = \{\phi \in \varPhi^*: \mathbf{v}_\phi \in K_i\}$ for $i \in \{1,\dots,8\}$.
Now each $\varPsi_i$ satisfies the condition in Lemma~\ref{lem-acyclic} and thus $|\varPsi_i| = O(n)$.
Since $\varPhi^* = \bigcup_{i=1}^8 \varPsi_i$, we have $|\varPhi^*| = O(n)$.
Therefore, we conclude the following.
\begin{lemma} \label{lem-linearcowcand}
    $|\varPhi(S,\mathcal{L}_C)| = O(n)$, where $n = |S|$.
\end{lemma}
Suppose $\varPhi(S,\mathcal{L}_C) = \{\phi_1,\dots,\phi_m\}$ where $m = O(n)$ and $\phi_1,\dots,\phi_m$ are sorted in increasing order of their lengths.
%By Lemma~\ref{lem-linearwedcand}, $m = O(n)$.
Now we only need a data structure which can report, for a query $C_q \in \mathcal{L}_C$, the smallest $i$ such that $\phi_i$ is contained in $C_q$.
Similar to the wedge case, we obtain such a data structure with $O(m)$ space and $O(\log m)$ query time (see \cite{xue2018searching}).
\begin{theorem} \label{thm-cowedds}
    There is an $O(n)$-space $\mathcal{L}_C$-RCP data structure with $O(\log n)$ query time.
\end{theorem}
Theorem~\ref{thm-plgres} now follows from Theorem~\ref{thm-reduction}, \ref{thm-wedds}, and \ref{thm-cowedds}.

\section{Translation RCP queries for smooth convex bodies} \label{sec-general}
Let $\varGamma$ be a fixed convex body whose boundary is \textit{smooth} (or \textit{smooth convex body}), that is, through each point on the boundary there is a unique tangent line to $\varGamma$.
%In this section, we study the $\mathcal{L}_\varGamma$-RCP problem when 
Assume we can compute in $O(1)$ time, for any line $l$ in $\mathbb{R}^2$, the segment $\varGamma \cap l$.
We investigate the $\mathcal{L}_\varGamma$-RCP problem (under the Euclidean metric).
Throughout this section, $O(\cdot)$ hides constants depending on $\varGamma$.
Our main result is the following, to prove which is the goal of this section.
\begin{theorem} \label{thm-general}
Let $\varGamma$ be a fixed smooth convex body in $\mathbb{R}^2$.
Then there is an $O(n \log n)$-space $\mathcal{L}_\varGamma$-RCP data structure with $O(\log^2 n)$ query time.
\end{theorem}
%For simplifying the exposition, in the rest of this section, we use the notation $C_p$ to denote the translation $p+C$ for a convex body $C$ in $\mathbb{R}^2$ and a point $p \in \mathbb{R}^2$.

\begin{figure}[h]
    \centering
    \begin{subfigure}[b]{5.5cm}
	    \includegraphics[height=1.8cm]{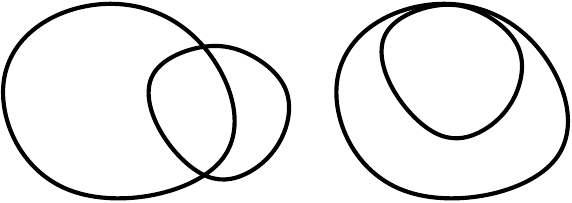}
	    \caption{Intersect plainly}
	\end{subfigure}	
    \hspace{1cm}
    \begin{subfigure}[b]{5.5cm}
	    \includegraphics[height=1.8cm]{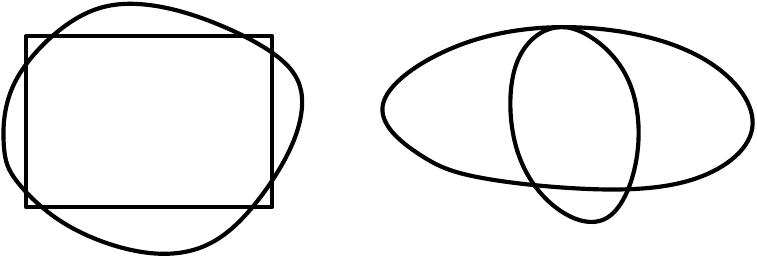}
	    \caption{Not intersect plainly}
	\end{subfigure}
    \caption{An illustration the concept of ``intersect plainly''}
    \label{figure:plainly intersect}
\end{figure}

\begin{figure}[b]
    \centering
    \includegraphics[height=4.5cm]{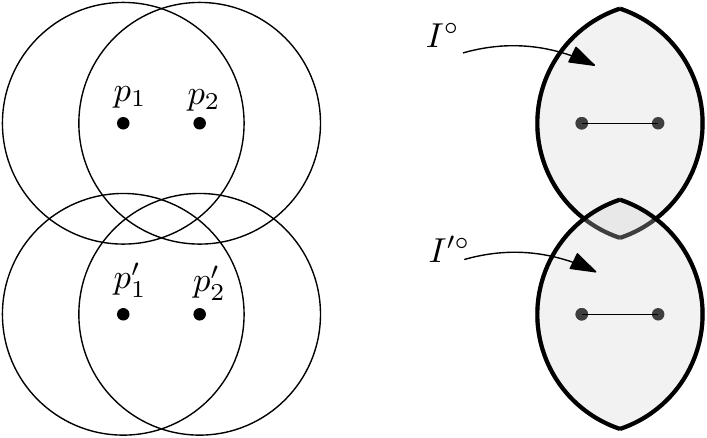}
    \caption{Illustration of Theorem~\ref{thm-4shapes} when $C$ is a disc}
    \label{fig-4shapes}
\end{figure}

%Suppose the dataset $S = \{a_1,\dots,a_n\}$ is given in $\mathbb{R}^2$.
%A pair $(a,b)$ with $a,b \in S$ is a \textit{candidate pair} if it is the closest-pair in some query range $q+X \in \mathcal{Q}$.
%The \textit{length} of a candidate pair $(a,b)$ is the length of the segment $[a,b]$.
Let $S$ be the given dataset in $\mathbb{R}^2$ of size $n$.
Suppose for convenience that the pairwise distances of the points in $S$ are distinct (so that the closest-pair in any subset of $S$ is unique).
Also, suppose that no three points in $S$ are collinear.
%To prove the theorem, our key step is to reduce the problem to the standard range-reporting problem (for $\varGamma$-translation queries).
Our data structure is based on the two technical results presented below, both of which are of geometric interest.
The first result states that sufficiently short candidate pairs do not cross when $\varGamma$ is a smooth convex body.
\begin{theorem} \label{thm-noncrossing}
	Let $\varGamma$ be a smooth convex body.
	Then there exists a constant $\tau > 0$ \textnormal{(}depending on $\varGamma$ only\textnormal{)} such that if $(a,b),(c,d) \in \varPhi(S,\mathcal{L}_\varGamma)$ are two pairs whose lengths are both at most $\tau$, then the segments $[a,b]$ and $[c,d]$ do not cross.
\end{theorem}
To introduce the second result, we need an important notion.
For two convex bodies $C,D$ in $\mathbb{R}^2$ such that $C \cap D \neq \emptyset$, we say $C$ and $D$ \textit{intersect plainly} if $\partial C \cap D$ and $\partial D \cap C$ are both connected; see Figure~\ref{figure:plainly intersect} for an illustration.
(The reader can intuitively understand this as ``the boundaries of $C$ and $D$ cross each other at most twice'', but it is in fact stronger.)
Note that a collection of convex bodies in $\mathbb{R}^2$ in which any two are disjoint or intersect plainly form a family of pseudo-discs \cite{agarwal2007shape_union}.
Our second result is the following theorem.
\begin{theorem} \label{thm-4shapes}
	Let $C$ be a convex body in $\mathbb{R}^2$, and $p_1,p_2,p_1',p_2' \in \mathbb{R}^2$ be four points \textnormal{(}not necessarily distinct\textnormal{)} such that $I^\circ \neq \emptyset$ and $I'^\circ \neq \emptyset$, where $I = C_{p_1} \cap C_{p_2}$ and $I' = C_{p_1'} \cap C_{p_2'}$.
	Suppose that any three of $p_1,p_2,p_1',p_2'$ are not collinear unless two of them coincide.
	If the segments $[p_1,p_2],[p_1',p_2']$ do not cross and $I \cap I' \neq \emptyset$, then $I$ and $I'$ intersect plainly.
\end{theorem}
Figure~\ref{fig-4shapes} gives an illustration of Theorem~\ref{thm-4shapes} in the case where $C$ is a disc.
Note that, even for the disc-case, without the condition that $[p_1,p_2],[p_1',p_2']$ do not cross, one can easily construct a counterexample in which $I$ and $I'$ do not intersect plainly.

In what follows, we first assume the correctness of the two theorems and present our $\mathcal{L}_\varGamma$-RCP data structure.
The proofs of the theorems will appear later in Section~\ref{sec-provingthms}.
%Let $\tau>0$ be the constant satisfying the condition in Theorem~\ref{thm-noncrossing}.
Our data structure consists of two parts $\mathcal{D}_1$ and $\mathcal{D}_2$ where $\mathcal{D}_1$ handles the queries for which the length of the answer (closest-pair) is ``short'' and $\mathcal{D}_2$ handles the queries for which the answer is ``long''.
%We use $\mathcal{N}$ to denote the collection of normal convex bodies in $\mathbb{R}^2$.
%\begin{theorem}
	%$\mathcal{N}$ contains all elliptical regions \textnormal{(}in particular, discs\textnormal{)} and all convex polygons with angles at least $\pi/2$ \textnormal{(}in particular, rectangles and regular $k$-polygons with $k \geq 4$\textnormal{)}.
%\end{theorem}

\subsection{Handling short-answer queries} \label{sec-reduction}
We describe the first part $\mathcal{D}_1$ of our data structure.
Let $\tau>0$ be the constant in Theorem~\ref{thm-noncrossing} such that any two candidate pairs of lengths at most $\tau$ do not cross.
For a query $\varGamma_q \in \mathcal{L}_\varGamma$, $\mathcal{D}_1$ reports the closest-pair in $S \cap \varGamma_q$ if its length is at most $\tau$, and reports nothing otherwise.

%Suppose $\varGamma$ is a smooth convex body, and a range-reporting data structure $\mathcal{D}_0$ for $\varGamma$-translation queries with $O(f(n))$ space and $O(g(n)+h(n) \cdot k)$ query time (where $k$ is the number of the points to be reported) is provided.
%With $\mathcal{D}_0$ in hand, we show how to construct an $\mathcal{L}_\varGamma$-RCP data structure with $O(f(n)+n \log n)$ space and $O(g(n)+h(n)+\log^2 n)$ query time.

%Applying Theorem~\ref{thm-noncrossing}, we find the constant $\tau>0$ such that any two candidate pairs of lengths at most $\tau$ do not cross.
Let $\varPhi_{\leq \tau} \subseteq \varPhi(S,\mathcal{L}_\varGamma)$ be the sub-collection consisting of the candidate pairs of lengths at most $\tau$, and suppose $m=|\varPhi_{\leq \tau}|$.
We have $m = O(n)$, because the graph $G = (S,\varPhi_{\leq \tau})$ is planar by Theorem~\ref{thm-noncrossing}.
Define $\tilde{\varGamma} = \{(x,y) \in \mathbb{R}^2: (-x,-y) \in \varGamma\}$.
For a pair $\theta = (a,b) \in \varPhi_{\leq \tau}$, we write $I^\theta = \tilde{\varGamma}_a \cap \tilde{\varGamma}_b$.
Then $\theta$ is contained in a query range $\varGamma_q \in \mathcal{L}_\varGamma$ iff $q \in I^\theta$.

In order to design $\mathcal{D}_1$, we first introduce a so-called \textit{membership} data structure (MDS).
Let $\varPsi = \{\theta_1,\dots,\theta_r\} \subseteq \varPhi_{\leq \tau}$ and $U = \bigcup_{\theta \in \varPsi} I^\theta$.
An MDS on $\varPsi$ can decide, for a given $\varGamma_q \in \mathcal{L}_\varGamma$, whether $\varGamma_q$ contains a pair in $\varPsi$ or not.
As argued before, $\varGamma_q$ contains a pair in $\varPsi$ iff $q \in U$.
Thus, to have an MDS on $\varPsi$, it suffices to have a data structure that can decide if a given point is in $U$.
By Theorem~\ref{thm-noncrossing}, the segments corresponding to any two pairs $\theta_i,\theta_j$ do not cross each other.
Also, no three points in $S$ are collinear by assumption.
Thus, by Theorem~\ref{thm-4shapes}, $I^{\theta_i}$ and $I^{\theta_j}$ intersect plainly for any $i,j \in \{1,\dots,r\}$ such that $I^{\theta_i} \cap I^{\theta_j} \neq \emptyset$.
It follows that $\{I^{\theta_1},\dots,I^{\theta_r}\}$ is a family of pseudo-discs \cite{agarwal2007shape_union}, and hence the complexity of their union $U$ is $O(r)$ by \cite{kedem1986union}.
As such, optimal point location data structures (e.g., \cite{edelsbrunner1986optimal,sarnak1986planar}) can be applied to decide whether a point is contained in $U$ in $O(\log r)$ time, using $O(r)$ space.
We remark that, although the edges defining the boundary of $U$ are not line-segments (existing point-location results we know of work on polygonal subdivisions), each edge is a connected portion of $\partial \tilde{\varGamma}$ and hence can be decomposed into constant number of ``fragments'' that are both $x$-monotone and $y$-monotone.
Recall our assumption that we can compute (in constant time) $\varGamma \cap l$ for any line $l$ in $\mathbb{R}^2$, and thus also $\tilde{\varGamma} \cap l$ for any line $l$.
%Specifically, we can compute the segment $\varGamma' \cap l$ for any line $l$ and the intersection $\varGamma'_p \cap \varGamma'_q$.
With this assumption, the existing data structures \cite{edelsbrunner1986optimal,sarnak1986planar} can be generalized straightforwardly for our purpose.
Thus, we have an MDS on $\varPsi$ with $O(r)$ space and $O(\log r)$ query time, which we denote by $\mathcal{M}(\varPsi)$.

With the MDS in hand, we can now design $\mathcal{D}_1$.
For a sub-collection $\varPsi = \{\theta_1,\dots,\theta_r\} \subseteq \varPhi_{\leq \tau}$ where $\theta_1,\dots,\theta_r$ are sorted in increasing order of their lengths, let $\mathcal{D}_1(\varPsi)$ be a data structure defined as follows.
If $r=1$, then $\mathcal{D}_1(\varPsi)$ simply stores the only pair $\theta_1 \in \varPsi$.
If $r>1$, let $\varPsi_1 = \{\theta_1,\dots,\theta_{\lfloor r/2 \rfloor}\}$ and $\varPsi_2 = \{\theta_{\lfloor r/2 \rfloor +1},\dots,\theta_r\}$.
Then $\mathcal{D}_1(\varPsi)$ consists of three parts: $\mathcal{D}_1(\varPsi_1)$, $\mathcal{D}_1(\varPsi_2)$, and $\mathcal{M}_{\varPsi_1}$, where $\mathcal{D}_1(\varPsi_1)$ and $\mathcal{D}_1(\varPsi_2)$ are defined recursively.
We show that we can use $\mathcal{D}_1(\varPsi)$ to find, for a query $\varGamma_q \in \mathcal{L}_\varGamma$, the shortest pair $\theta^* \in \varPsi$ contained in $\varGamma_q$.
We first query $\mathcal{M}(\varPsi_1)$ to see if $\varGamma_q$ contains a pair in $\varPsi_1$.
If so, $\theta^*$ must be in $\varPsi_1$, so we recursively query $\mathcal{D}_1(\varPsi_1)$ to find it.
Otherwise, we recursively query $\mathcal{D}_1(\varPsi_2)$.
In this way, we can eventually find $\theta^*$.
Now we simply define $\mathcal{D}_1 = \mathcal{D}_1(\varPhi_{\leq \tau})$.

The space and query time of $\mathcal{D}_1$ can be bounded by a direct analysis.
%We follow the notations used in Section~\ref{sec-reduction}.
Define $\mathsf{sp}(r)$ and $\mathsf{qt}(r)$ as the maximum space and query time of $\mathcal{D}_1(\varPsi)$ for any sub-collection $\varPsi \subseteq \varPhi_{\leq \tau}$ with $|\varPsi| = r$, respectively.
We show that $\mathsf{sp}(r) = O(r \log r)$ and $\mathsf{qt}(r) = O(\log^2 r)$.
%For a sub-collection $\varPsi \subseteq \varPhi_{\leq \tau}$, we show that the space of $\mathcal{D}_1(\varPsi)$ is $O(r \log r)$ and the query time of $\mathcal{D}_1(\varPsi)$ is $O(\log^2 r)$, where $r = |\varPsi|$.
Let $\varPsi = \{\theta_1,\dots,\theta_r\} \subseteq \varPhi_{\leq \tau}$ where $\theta_1,\dots,\theta_m$ are sorted in increasing order of their lengths.
Write $\varPsi_1 = \{\theta_1,\dots,\theta_{\lfloor r/2 \rfloor}\}$ and $\varPsi_2 = \{\theta_{\lfloor r/2 \rfloor+1},\dots,\theta_m\}$.
Recall that $\mathcal{D}_1(\varPsi)$ consists of three parts: $\mathcal{D}_1(\varPsi_1)$, $\mathcal{D}_1(\varPsi_2)$, and $\mathcal{M}(\varPsi_1)$.
We know that the space of $\mathcal{M}(\varPsi_1)$ is $O(r)$, which implies the recurrence
\begin{equation*}
    \mathsf{sp}(r) \leq 2 \cdot \mathsf{sp}(r/2) + O(r).
\end{equation*}
The above recurrence solves to $\mathsf{sp}(r) = O(r \log r)$.
To analyze the query time, we recall the query process of $\mathcal{D}_1(\varPsi)$.
When querying $\mathcal{D}_1(\varPsi)$, we first query $\mathcal{M}(\varPsi_1)$ and then query either $\mathcal{D}_1(\varPsi_1)$ or $\mathcal{D}_1(\varPsi_2)$.
We know that the query time of $\mathcal{M}(\varPsi_1)$ is $O(\log r)$, which implies the recurrence
\begin{equation*}
    \mathsf{qt}(r) \leq \mathsf{qt}(r/2) + O(\log r).
\end{equation*}
The above recurrence solves to $\mathsf{qt}(r) = O(\log^2 r)$.
Since $\mathcal{D}_1 = \mathcal{D}_1(\varPhi_{\leq \tau})$, we conclude that the space of $\mathcal{D}_1$ is $O(m \log m)$ and the query time of $\mathcal{D}_1$ is $O(\log^2 m)$ where $m = |\varPhi_{\leq \tau}|$.

\subsection{Handling long-answer queries} \label{sec-long}
If $\mathcal{D}_1$ fails to answer the query $\varGamma_q$, then the length of the closest-pair in $S \cap \varGamma_q$ is greater than $\tau$.
In this case, we shall use the second part $\mathcal{D}_2$ of our data structure to answer the query.
$\mathcal{D}_2$ simply reports all the points in $S \cap \varGamma_q$ and computes the closest-pair by brute-force.
Since the length of the closest-pair in $S \cap \varGamma_q$ is greater than $\tau$, we have $|S \cap \varGamma_q| = O(1)$ by Lemma~\ref{lem-pigeon} and hence computing the closest-pair takes $O(1)$ time.
In order to do reporting, we consider the problem in the dual setting.
Again, define $\tilde{\varGamma} = \{(x,y) \in \mathbb{R}^2: (-x,-y) \in \varGamma\}$.
Clearly, for any $a \in \mathbb{R}^2$, $a \in \varGamma_q$ iff $q \in \tilde{\varGamma}_a$.
Thus, the problem is equivalent to reporting the ranges in $\mathcal{S} = \{\tilde{\varGamma}_a: a \in S\}$ that contain $q$.
Define the \textit{depth}, $\text{dep}(p)$, of a point $p \in \mathbb{R}^2$ as the number of the ranges in $\mathcal{S}$ containing $p$.
Let $\mathcal{A}$ be the arrangement of the ranges in $\mathcal{S}$, and $k$ be a sufficiently large constant.
The $\leq k$-level of $\mathcal{A}$, denoted by $\mathcal{A}_{\leq k}$, is the sub-arrangement of $\mathcal{A}$ contained in the region $R_{\leq k} = \{p \in \mathbb{R}^2:\text{dep}(p) \leq k\}$.
By Theorem~\ref{thm-4shapes}, any two ranges $\tilde{\varGamma}_a,\tilde{\varGamma}_b \in \mathcal{S}$ intersect plainly if they intersect (setting $p_1 = p_2 = a$ and $p_1' = p_2' = b$ when applying Theorem~\ref{thm-4shapes}), which implies that $\mathcal{S}$ is a family of $n$ pseudo-discs and $\mathcal{A}$ is a pseudo-disc arrangement.
So we have the following well-known lemma.
\begin{lemma} \label{lem-level}
    The complexity of $\mathcal{A}_{\leq k}$ is $O(n)$ for a constant $k$.
\end{lemma}
By the above lemma, we can build a point-location data structure on $\mathcal{A}_{\leq k}$ with $O(n)$ space and $O(\log n)$ query time.
%(As argued in Section~\ref{sec-reduction}, although the edges of $\mathcal{A}_{\leq k}$ are not line-segments, the existing point-location data structures \cite{edelsbrunner1986optimal,sarnak1986planar} still apply.)
Also, we associate to each cell $\Delta$ of $\mathcal{A}_{\leq k}$ the (at most $k$) ranges in $\mathcal{S}$ containing $\Delta$.
Now we can report the ranges in $\mathcal{S}$ containing $q$ as follows.
Since $|S \cap \varGamma_q| = O(1)$ and $k$ is sufficiently large, we have $|S \cap \varGamma_q| \leq k$ and hence $q$ is in $\mathcal{A}_{\leq k}$.
Using the point-location data structure, we find in $O(\log n)$ time the cell $\Delta$ of $\mathcal{A}_{\leq k}$ containing $q$.
Then the ranges associated to $\Delta$ are exactly those containing $q$.
Together with our argument above, this gives us the desired data structure $\mathcal{D}_2$ with $O(n)$ space and $O(\log n)$ time.
Combining $\mathcal{D}_2$ with the data structure $\mathcal{D}_1$ in the last section, Theorem~\ref{thm-general} is proved.

\subsection{Proving the technical theorems} \label{sec-provingthms}
\begin{figure}[h]
	\centering
	\begin{subfigure}[t]{5.5cm}
	    \includegraphics[height=3.3cm]{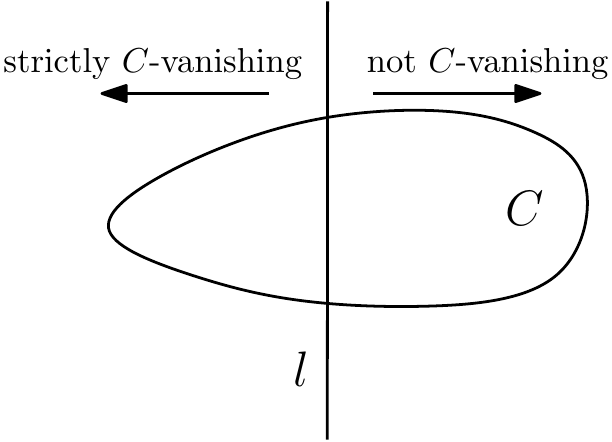}
	    \caption{A line with a strictly $C$-vanishing side and a side that is not $C$-vanishing}
	\end{subfigure}	
    \hspace{1cm}
	\begin{subfigure}[t]{5.5cm}
	    \includegraphics[height=3.3cm]{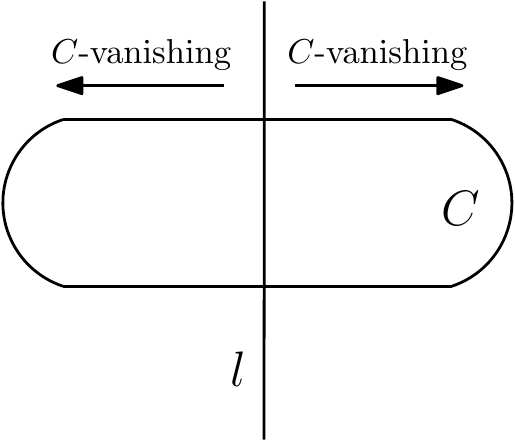}
	    \caption{A line with two $C$-vanishing sides}
	\end{subfigure}	
    \caption{An illustration of ``$C$-vanishing''}
    \label{fig-vanishing}
\end{figure}

This section is dedicated to prove Theorem~\ref{thm-noncrossing} and~\ref{thm-4shapes}.
To this end, we first introduce some basic notions and geometric results regarding convex bodies in $\mathbb{R}^2$.
Let $C$ be a convex body in $\mathbb{R}^2$.
%For a point $p \in \mathbb{R}^2$, we denote by $C_p$ the translation $p+C$.
For a line $l$ in $\mathbb{R}^2$, we denote by $\text{len}_C(l)$ the length of the segment $C \cap l$.
Suppose $l$ is given by the equation $ax+by+c=0$, then it cuts $\mathbb{R}^2$ into two (closed) halfplanes, $ax+by+c \geq 0$ and $ax+by+c \leq 0$ (we call them the two \textit{sides} of $l$ hereafter).
Let $H$ be one side of $l$.
For a real number $t \geq 0$, define $l_t$ as the (unique) line parallel to $l$ satisfying $l_t \subseteq H$ and $\text{dist}(l,l_t) = t$.
Suppose $\lambda = \sup \{t \geq 0: C \cap l_t \neq \emptyset\}$ (if $\{t \geq 0: C \cap l_t \neq \emptyset\} = \emptyset$, set $\lambda = 0$).
\begin{definition}
	We say $H$ is a $C$-\textbf{vanishing} \textnormal{(}resp., \textbf{strictly} $C$-\textbf{vanishing}\textnormal{)} side of $l$, if $f(t) = \textnormal{len}_C(l_t)$ is a monotonically non-increasing \textnormal{(}resp., decreasing\textnormal{)} function in the domain $[0,\lambda]$.
\end{definition}
%Moreover, if $f(t) = \text{len}_C(l_t)$ is a monotonically decreasing (resp., increasing) function in the domain $t \geq 0$ (resp., $t \leq 0$), we say $H_+$ (resp., $H_-$) is a \textit{strictly} $C$-\textit{vanishing} side of $l$.
Note that at least one side of $l$ is $C$-vanishing due to the convexity of $C$.
Furthermore, if $\text{len}_C(l) \geq \text{len}_C(l')$ for any line $l'$ parallel to $l$, then the two sides of $l$ are both $C$-vanishing, otherwise one side of $l$ is strictly $C$-vanishing while the other one is not $C$-vanishing; see Figure~\ref{fig-vanishing} for an illustration.
In particular, for any line in $\mathbb{R}^2$, either its two sides are both $C$-vanishing or it has a strictly $C$-vanishing side (this observation will be used later in the proofs of the theorems).
With the above definition, we can introduce our first lemma regarding the intersection of two $\varGamma$-translates.
%\ifdefined\FullPage
%\fi
%The following lemma describes another basic property of a $C$-vanishing side of a line.

%Note that if $C$ and $D$ plainly intersect, then $\partial C \cap D^\circ$ and $\partial D \cap C^\circ$ should also be connected, as one can easily verify that $\partial C \cap D^\circ = \partial I \backslash (\partial D \cap C)$ and $\partial D \cap C^\circ = \partial I \backslash (\partial C \cap D)$, where $I = C \cap D$.
%The following lemma basically claims that two translations of the same convex body plainly intersect (if they intersect).
\begin{lemma} \label{lem-2shapes}
	Let $C$ be a convex body in $\mathbb{R}^2$ and $p_1,p_2 \in \mathbb{R}^2$ be two points.
	Also, let $I = C_{p_1} \cap C_{p_2}$.
	If $I \neq \emptyset$, then $C_{p_1}$ and $C_{p_2}$ plainly intersect.
	In addition, if $I^\circ \neq \emptyset$, then there exist two distinct points $u,v \in \partial I$ and the line $l$ through $u,v$ such that \\
	\textnormal{(1)} the two sides of $l$ are both $I$-vanishing; \\
	\textnormal{(2)} $l$ is not parallel to the line through $p_1,p_2$; \\
	\textnormal{(3)} $u,v$ are the endpoints of the segment $I \cap l$; \\
	\textnormal{(4)} the two arcs in $\partial I$ connecting $u,v$ are contained in $\partial C_{p_1}$ and $\partial C_{p_2}$ respectively.
	%Furthermore, there exists a line $l$ with two $I$-vanishing sides $H_1$ and $H_2$ such that for each $i \in \{1,2\}$ either $I \cap H_i = C_{p_i} \cap H_i$ or $I \cap H_i = I \cap l$, where $I = C_{p_1} \cap C_{p_2}$.
\end{lemma}
An immediate corollary of the above lemma is presented below.
\begin{corollary} \label{cor-halfcontain}
	Let $C,p_1,p_2$ be as in Lemma~\ref{lem-2shapes}, and $l$ be an arbitrary line in $\mathbb{R}^2$. \\
	\textnormal{(1)} Suppose $C_{p_1} \cap l \subsetneq C_{p_2} \cap l$ and $V$ is a $C_{p_1}$-vanishing side of $l$.
	Then $C_{p_1} \cap V \subseteq C_{p_2}$. \\
	\textnormal{(2)} Suppose $C_{p_1} \cap l$ is in the ``interior'' of $C_{p_2} \cap l$ \textnormal{(}i.e., $C_{p_2} \cap l$ excluding both endpoints\textnormal{)} and $V$ is a $C_{p_1}$-vanishing side of $l$. 
	Then $C_{p_1} \cap (V \backslash l) \subseteq C_{p_2} \backslash \partial C_{p_2}$.
	%\textnormal{(2)} Suppose $C_{p_1} \cap C_{p_2} \cap l = \emptyset$. 
	%Then either $C_{p_1} \cap C_{p_2} \cap H_1 = \emptyset$ or $C_{p_1} \cap C_{p_2} \cap H_2 = \emptyset$, where $H_1,H_2$ are the two sides of $l$.
	%$l$ be a line such that $C_{p_1} \cap l \subsetneq C_{p_2} \cap l$.
	%If $V$ is a $C_{p_1}$-vanishing side of $l$, then $C_{p_1} \cap V \subseteq C_{p_2}$.
\end{corollary}
The following two lemmas will be used frequently in the proof of Theorem~\ref{thm-4shapes}.
\begin{lemma} \label{lem-contain}
	Let $C,p_1,p_2$ be as in Lemma~\ref{lem-2shapes} such that $I^\circ \neq \emptyset$ for $I = C_{p_1} \cap C_{p_2}$.
	Also, let $u,v,l$ be the points and line satisfying the conditions in Lemma~\ref{lem-2shapes}.
	For any point $p \in \mathbb{R}^2$ such that $p,p_1,p_2$ are not collinear, if $I \cap l \subseteq C_p$ \textnormal{(}resp., $I \cap l \subseteq C_p^\circ$\textnormal{)}, then $I \subseteq C_p$ \textnormal{(}resp., $I \subseteq C_p^\circ$\textnormal{)}.
\end{lemma}
\begin{lemma} \label{lem-exclude}
	Let $C,p_1,p_2$ be as in Lemma~\ref{lem-2shapes} such that $I^\circ \neq \emptyset$ for $I = C_{p_1} \cap C_{p_2}$.
	Also, let $u,v,l$ be the points and line satisfying the conditions in Lemma~\ref{lem-2shapes}.
	For any point $p \in \mathbb{R}^2$ such that $p,p_1,p_2$ are not collinear, if $u,v \notin C_p^\circ$ and $I^\circ \cap C_p^\circ \neq \emptyset$, then $C_p \cap I = C_p \cap C_{p_i}$ for some $i \in \{1,2\}$.
\end{lemma}

\subsubsection{Proof of Theorem~\ref{thm-noncrossing}} \label{sec-prf1}
In order to prove Theorem~\ref{thm-noncrossing}, we need the following key observation.
\begin{lemma} \label{lem-tau}
	Let $C$ be a smooth convex body in $\mathbb{R}^2$.
	Then there exists a number $\tau>0$ satisfying the following condition.
	For any line $l$ with $0 < \textnormal{len}_C(l) < \tau$ and any point $r \in C$ on a $C$-vanishing side of $l$, the distance between $r$ and an \textnormal{(}arbitrary\textnormal{)} endpoint of $C \cap l$ is less than $\textnormal{len}_C(l)$.
\end{lemma}

\begin{figure}[htbp]
    \begin{center}
        \includegraphics[height=4cm]{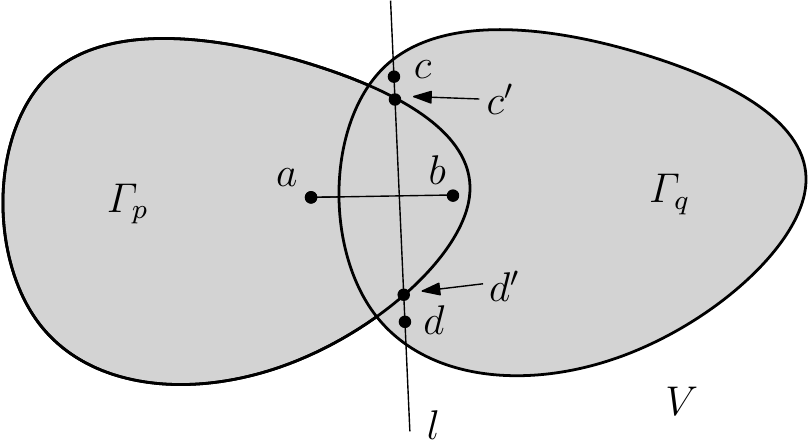}
    \end{center}
    \caption{Illustrating the proof of Theorem~\ref{thm-noncrossing}.}
    \label{fig-thmnoncross}
\end{figure}

Now we are able to prove Theorem~\ref{thm-noncrossing}.
Suppose $\varGamma$ is a smooth convex body in $\mathbb{R}^2$.
Taking $C=\varGamma$, we can find a constant $\tau$ satisfying the condition in Lemma~\ref{lem-tau}.
We claim that $\tau$ also satisfies the condition in Theorem~\ref{thm-noncrossing}.
Suppose $(a,b)$ and $(c,d)$ are two candidate pairs of lengths at most $\tau$.
Assume that $[a,b]$ and $[c,d]$ cross.
By Lemma~\ref{lem-cross}, there exists $\varGamma_p \in \mathcal{L}_\varGamma$ such that $\varGamma_p \cap \{a,b,c,d\} = \{a,b\}$ or $\varGamma_p \cap \{a,b,c,d\} = \{c,d\}$.
Assume $\varGamma_p \cap \{a,b,c,d\} = \{a,b\}$.
Let $q \in \mathbb{R}^2$ be a point such that the closest-pair in $\varGamma_q$ is $(c,d)$; then $c,d \in \varGamma_q$.
%Let $q_1,q_2 \in \mathbb{R}^2$ be two points such that the closest-pairs in $X_{q_1}$ and $X_{q_2}$ are $(a,b)$ and $(c,d)$ respectively.
%Then $a,b \in X_{q_1}$ and $c,d \in X_{q_2}$.
%If $X_{q_1} \cap \{c,d\} \neq \emptyset$ and $X_{q_2} \cap \{a,b\} \neq \emptyset$, then we can apply the argument in the proof of Lemma~\ref{theorem:wedge arrangement is planar} to obtain a contradiction.
%Therefore, we have either $c,d \notin X_{q_1}$ or $a,b \notin X_{q_2}$.
%Without loss of generality, assume $c,d \notin X_{q_1}$.
Let $l$ be the line through $c,d$, and $V$ be a $\varGamma_p$-vanishing side of $l$.
See Figure~\ref{fig-thmnoncross} for an illustration of the notations.
Since the segments $[a,b]$ and $[c,d]$ cross, $[c,d]$ must intersect $\varGamma_p$.
But $c,d \notin \varGamma_p$, thus $\varGamma_p \cap l \subsetneq [c,d]$.
Furthermore, because $[c,d] \subseteq \varGamma_q \cap l$, we have $\varGamma_p \cap l \subsetneq \varGamma_q \cap l$.
This implies $\varGamma_p \cap V \subseteq \varGamma_q$ by (1) of Corollary~\ref{cor-halfcontain}.
Note that one of $a,b$ must be contained in $\varGamma_p \cap V$, say $b \in \varGamma_p \cap V$.
Thus $b \in \varGamma_q$.
We now show that $\text{dist}(b,c) < \text{dist}(c,d)$ and $\text{dist}(b,d) < \text{dist}(c,d)$.
As $\varGamma_p \cap l \subsetneq [c,d]$ and the length of $[c,d]$ is at most $\tau$, we have $\text{len}_{\varGamma_p}(l) < \tau$.
We denote by $c',d'$ the two endpoints of the segment $\varGamma_p \cap l$; see Figure~\ref{fig-thmnoncross}.
By Lemma~\ref{lem-tau}, the distance from $c'$ (or $d'$) to any point in $\varGamma_p \cap V$ is less than $\text{len}_{\varGamma_p}(l) = \text{dist}(c',d')$.
In particular, $\text{dist}(b,c') < \text{dist}(c',d')$ and $\text{dist}(b,d') < \text{dist}(c',d')$.
Now $\text{dist}(b,c) \leq \text{dist}(b,c') + \text{dist}(c',c) < \text{dist}(c',d') + \text{dist}(c',c) = \text{dist}(c,d') < \text{dist}(c,d)$.
For the same reason, we have $\text{dist}(b,d) < \text{dist}(c,d)$.
Since $b,c,d \in \varGamma_q$, this contradicts the fact that $(c,d)$ is the closest-pair in $\varGamma_q$.
The proof of Theorem~\ref{thm-noncrossing} is now complete.

\subsubsection{Proof of Theorem~\ref{thm-4shapes}} \label{sec-prf2}
We first prove some special cases of Theorem~\ref{thm-4shapes}.
The conclusions will be used in the final proof of Theorem~\ref{thm-4shapes}.
The following lemma handles the special case $p_1' = p_2'$.
In this case, $I'$ is a translate of $C$.
\begin{lemma} \label{lem-3shapes}
	Let $C,p_1,p_2$ be as in Lemma~\ref{lem-2shapes} such that $I^\circ \neq \emptyset$ for $I = C_{p_1} \cap C_{p_2}$.
	For any point $p \in \mathbb{R}^2$ such that $p,p_1,p_2$ are not collinear, if $I \cap C_p \neq \emptyset$, then $I$ and $C_p$ plainly intersect.
\end{lemma}
The next lemma considers another special case of Theorem~\ref{thm-4shapes}.
\begin{lemma} \label{lem-interior}
	Let $C,p_1,p_2$ be as in Lemma~\ref{lem-2shapes} such that $I^\circ \neq \emptyset$ for $I = C_{p_1} \cap C_{p_2}$, and $u,v$ be the points satisfying the conditions in Lemma~\ref{lem-2shapes}.
	Also, let $p_1',p_2' \in \mathbb{R}^2$ be two points such that $I'^\circ \neq \emptyset$ for $I' = C_{p_1'} \cap C_{p_2'}$.
	If $u \in I'^\circ$ or $v \in I'^\circ$, then $I$ and $I'$ plainly intersect.
\end{lemma}
The last ingredient needed for proving Theorem~\ref{thm-4shapes} is the following observation.
%As we will see, it explains why the condition that $[p_1,p_2]$ and $[p_1',p_2']$ do not cross plays an important role in the theorem.
\begin{lemma} \label{lem-topbottom}
	Let $C,p_1,p_2$ be as in Lemma~\ref{lem-2shapes} such that $I^\circ \neq \emptyset$ for $I = C_{p_1} \cap C_{p_2}$.
	Suppose $p_1,p_2$ are on the $x$-axis.
	Also, let $u,v$ be the points satisfying the conditions in Lemma~\ref{lem-2shapes}.
	By \textnormal{(2)} of Lemma~\ref{lem-2shapes}, the line through $u,v$ is not horizontal and thus $u.y \neq v.y$.
	Assume $u.y > v.y$.
	If $p \in \mathbb{R}^2$ is a point with $p.y>0$, then $v \in C_p$ \textnormal{(}resp., $v \in C_p^\circ$\textnormal{)} only if $u \in C_p$ \textnormal{(}resp., $u \in C_p^\circ$\textnormal{)}.
\end{lemma}
\ifdefined\SODA
    \vspace{-0.1cm}
\fi

Now we are able to prove Theorem~\ref{thm-4shapes}.
If $I^\circ \cap I'^\circ = \emptyset$, it is easy to verify that $\partial I \cap I' = \partial I' \cap I = I \cap I'$, which is connected.
So assume $I^\circ \cap I'^\circ \neq \emptyset$.
We consider two cases separately.
In the first case, there exist $i,j \in \{1,2\}$ such that $p_i = p_j'$.
The second case is the complement of the first one.

To handle the first case is easy.
Without loss of generality, we may assume $p_1 = p_1'$.
Then $\partial I \cap I' = \partial I \cap C_{p_2'}$.
If $p_2' = p_2$ or $p_2' = p_1$, then $\partial I \cap C_{p_2'} = \partial I$, which is clearly connected.
If $p_1 = p_2$, then $\partial I \cap C_{p_2'} = \partial C_{p_1} \cap C_{p_2'}$, which is connected by Lemma~\ref{lem-2shapes}.
Otherwise, $p_1,p_2,p_2'$ are distinct and hence not collinear by assumption.
In this situation, Lemma~\ref{lem-3shapes} implies the connectedness of $\partial I \cap C_{p_2'}$.
Symmetrically, $\partial I' \cap I$ is also connected.
Therefore, $I$ and $I'$ plainly intersect.

The second case is subtler.
In this case, if $p_1 = p_2$ or $p_1' = p_2'$, Lemma~\ref{lem-2shapes} and Lemma~\ref{lem-3shapes} apply to show that $I$ and $I'$ plainly intersect.
So suppose $p_1,p_2,p_1',p_2'$ are distinct.
Then any three points are not collinear by assumption.
Since the segments $[p_1,p_2]$ and $[p_1',p_2']$ do not cross, we may assume that $p_1',p_2'$ are on the same side of the line through $p_1,p_2$.
Without loss of generality, assume that $p_1,p_2$ are both on the $x$-axis and $p_1',p_2'$ are above the $x$-axis ($p_1'.y>0$ and $p_2'.y>0$).
Let $u,v,l$ be the points and line satisfying the conditions in Lemma~\ref{lem-2shapes}, and $u',v',l'$ be the counterparts of $u,v,l$ for the convex bodis $C_{p_1'},C_{p_2'}$.
By (1) and (4) of Lemma~\ref{lem-2shapes}, $l$ has two $I$-vanishing sides (say $H_1$ and $H_2$), and the two arcs in $\partial I$ connecting $u,v$ are contained in $\partial C_{p_1}$ and $\partial C_{p_2}$ respectively (we denote by $\sigma_i$ the arc contained in $\partial C_{p_i}$ for $i \in \{1,2\}$ and assume $\sigma_i \subseteq H_i$ for $i \in \{1,2\}$ without loss of generality).
Moreover, by (2) of Lemma~\ref{lem-2shapes}, $l$ is not horizontal and thus $u.y \neq v.y$.
Assume that $u.y > v.y$.
First, we prove the connectedness of $\partial I' \cap I$.
If $u \in C_{p_i'}^\circ$ for all $i \in \{1,2\}$, then $u \in I'^\circ$ and Lemma~\ref{lem-interior} immediately applies to show that $I$ and $I'$ plainly intersect.
Otherwise $u \notin C_{p_i'}^\circ$ for some $i \in \{1,2\}$, say $u \notin C_{p_1'}^\circ$.
By Lemma~\ref{lem-topbottom}, we have $v \notin C_{p_1'}^\circ$.
In this situation, Lemma~\ref{lem-exclude} implies that either $I \cap C_{p_1'} = C_{p_1} \cap C_{p_1'}$ or $I \cap C_{p_1'} = C_{p_2} \cap C_{p_1'}$ (recall that we already assumed $I^\circ \cap I'^\circ \neq \emptyset$).
It follows that either $\partial I' \cap I = \partial I' \cap C_{p_1}$ or $\partial I' \cap I = \partial I' \cap C_{p_2}$ (since $\partial I' \subseteq C_{p_1'}$).
We can then apply Lemma~\ref{lem-3shapes} to deduce the connectedness of $\partial I' \cap I$.
Next,  we prove the connectedness of $\partial I \cap I'$.
If $v \in C_{p_1'}$ (resp., $v \in C_{p_2'}$), then by Lemma~\ref{lem-topbottom} we have $u \in C_{p_1'}$ (resp., $u \in C_{p_2'}$).
In this situation, Lemma~\ref{lem-contain} applies to show that $I \subseteq C_{p_1'}$ (resp., $I \subseteq C_{p_2'}$) and $\partial I \cap I' = \partial I \cap C_{p_2'}$ (resp., $\partial I \cap I' = \partial I \cap C_{p_1'}$).
Then Lemma~\ref{lem-3shapes} immediately implies the connectedness of $\partial I \cap I'$.
The remaining case is that $v \notin C_{p_i'}$ for all $i \in \{1,2\}$.
We observe that $\partial I \cap I' = (\partial I \cap C_{p_1'}) \cap (\partial I \cap C_{p_2'})$.
Now both $\partial I \cap C_{p_1'}$ and $\partial I \cap C_{p_2'}$ are connected by Lemma~\ref{lem-3shapes}.
In addition, they are both subsets of $\partial I \backslash \{v\}$.
But $\partial I \backslash \{v\}$ is homeomorphic to the real line $\mathbb{R}$ and the intersection of any connected subsets of $\mathbb{R}$ is connected.
Therefore, $\partial I \cap I'$ is connected.
The proof of Theorem~\ref{thm-4shapes} is now complete.

\bibliography{my_bib}

%\newpage
\appendix
%\noindent
%\textbf{\LARGE Appendix}
\section{Missing proofs} \label{appx-proof}

\subsection{Proof of Lemma~\ref{lem-pigeon}}
Let $\alpha$ be the diameter of $\varGamma$, i.e., $\alpha = \sup_{a,b \in \varGamma} \text{dist}(a,b)$.
Then we can find a square $Z$ of side-length $\alpha$ such that $\varGamma \subseteq Z$.
Define $c = \lceil \alpha/\mu \rceil$.
%Without loss of generality, assume $\alpha = c \delta$ for a positive integer $c$.
Suppose $p \in \mathbb{R}^2$ is a point such that the closest-pair in $S \cap \varGamma_p$ has length greater than or equal to $\mu$.
We must show that $|S \cap \varGamma_p|$ is bounded by some constant.
Note that $S \cap \varGamma_p \subseteq Z_p$.
We evenly decompose $Z_p$ into $2c \times 2c$ small squares of side-length $\alpha/(2c) \leq \mu/2$.
Since the closest-pair in $S \cap \varGamma_p$ has length greater than or equal to $\mu$, the distance between any two distinct points in $S \cap \varGamma_p$ is at least $\mu$.
Therefore, each small square contains at most one point in $S \cap \varGamma_p$.
By the Pigeonhole Principle, $|S \cap \varGamma_p| \leq 4c^2$.
This completes the proof as $c$ is a constant.

\subsection{Proof of Lemma~\ref{lem-cross}}
We prove the lemma by contradiction.
Assume that no $X \in \mathcal{X}$ satisfies the desired property.
Let $X,X' \in \mathcal{X}$ be the two ranges in which the closest-pairs are $(a,b)$ and $(a',b')$, respectively.
Then, $a,b \in X$ and $a',b' \in X'$.
By our assumption, $X$ contains at least one of $a'$ and $b'$, while $X'$ contains at least one of $a$ and $b$.
Without loss of generality, assume $a' \in X$ and $a \in X'$.
Let $c$ be the intersection point of the segments $[a,b]$ and $[a',b']$.
Since $(a,b)$ is the closest-pair in $X$ and $a' \in X$, $\text{dist}(a,b) \leq \text{dist}(a',b)$.
For the same reason, $\text{dist}(a',b') \leq \text{dist}(a,b')$.
It follows that $\text{dist}(a,b) + \text{dist}(a',b') \leq \text{dist}(a',b) + \text{dist}(a,b')$.
%Now, for a contradiction, assume $[a, b]$ and $[c, d]$ properly intersect at $e$.
On the other hand, by the triangle inequality, $\text{dist}(a,c) + \text{dist}(b',c) > \text{dist}(a,b')$ and $\text{dist}(a',c) + \text{dist}(b,c) > \text{dist}(a',b)$.
This implies $\text{dist}(a,b) + \text{dist}(a',b') = \text{dist}(a,c) + \text{dist}(b,c) + \text{dist}(a',c) + \text{dist}(b',c) > \text{dist}(a,b') + \text{dist}(a',b)$, which results in a contradiction.

\subsection{Proof of Lemma~\ref{lem-rangerep}}
For a cell $\Box$ of $G$, let $m_\Box$ be the number of the points in $S_\Box$.
First, we notice that there are $O(n)$ cells $\Box$ of $G$ such that $m_\Box>0$; we call them \textit{nonempty} cells.
For each nonempty cell $\Box$ and each $W \in \mathcal{W}_\varGamma$, we build a range-reporting data structure on $S_\Box$ for $W$-translation queries.
There exists such a data structure with $O(m_\Box)$ space and $O(\log m_\Box + k)$ query time (where $k$ is the number of the reported points), no matter whether $W$ is a wedge or a co-wedge.
Indeed, if $W$ is a wedge, we can simply apply an affine transformation to the dataset so that a $W$-translation range-reporting query on the original dataset is equivalent to a quadrant range-reporting query (or dominance range-reporting query) on the new dataset.
For example, if $W = \{\alpha x+\beta y + \gamma \geq 0\} \cap \{\alpha' x+\beta' y + \gamma' \geq 0\}$, we can apply the affine transformation $f:(x,y) \mapsto (\alpha x+\beta y + \gamma, \alpha' x+\beta' y + \gamma')$ to the dataset $S$ and then a $W$-translation range-reporting query on $S$ is equivalent to a northeast quadrant range-reporting query on $f(S)$.
Note that the quadrant range-reporting can be solved optimally using, for example, priority search trees \cite{de1997computational}, which gives us an optimal range-reporting data structure on $S_\Box$ for $W$-translation queries.
If $W$ is a co-wedge, it is the union of two halfplanes.
By the decomposability of range reporting, it suffices to do range-reporting for halfplane translation queries, which can clearly be solved optimally.
Since $\sum m_\Box = n$, the overall space cost of these data structures is $O(n)$.

To answer a query $\varGamma_q \in \mathcal{L}_\varGamma$, we first find all cells of $G$ that intersect $\varGamma_q$.
The number of these cells is $O(1)$, as it can be upper bounded by $\Delta^2/(\delta/2)^2$ where $\Delta$ is the diameter of $\varGamma$.
Finding these cells can be easily done in $O(\log n)$ time; see Appendix~\ref{appx-cell}.
For each such cell $\Box$, we find the (co-)wedge $W \in \mathcal{W}_\varGamma$ such that $\Box \cap \varGamma_q = \Box \cap W_q$ and query the corresponding associated data structure to report the points in $S_\Box \cap \varGamma_q$, which can be done in $O(\log m_\Box + k_\Box)$ time, where $k_\Box$ is the number of the reported points.
Since $\sum k_\Box = k$, the overall query time is $O(\log n + k)$.

\subsection{Proof of Lemma~\ref{lem-qcell}}
We first notice that for two points $a,b \in \mathbb{R}^2$ satisfying $\text{dist}(a,b) \leq \delta/4$, the segment $[a,b]$ crosses at most one horizontal line and one vertical line of $G$, and thus there must exist a quad-cell $\boxplus$ of $G$ such that $a,b \in \boxplus$.
If the length of $\phi^*$ is at most $\delta/4$, the length of the closest-pair in $S \cap \varGamma_q$ is also at most $\delta/4$, which implies that the closest-pair is contained in some quad-cell of $G$, and hence it must be $\phi^*$.
On the other hand, if $\phi^*$ has a length greater than $\delta/4$, then either the closest-pair in $S \cap \varGamma_q$ is $\phi^*$ or it is not contained in any quad-cell of $G$.
In either of the two cases, the length of the closest-pair in $S \cap \varGamma_q$ is greater than $\delta/4$, and we have $|S \cap \varGamma_q| = O(1)$ by Lemma~\ref{lem-pigeon}.

\subsection{Proof of Lemma~\ref{lem-smallest}}
We first prove the existence of the smallest $W$-translate containing $A$.
%If $|A|=1$, we are done; indeed, if $|A| = \{a\}$, then $W_a$ is just the unique $W$-translate containing $A$.
%Assume for any set $A \subseteq \mathbb{R}^2$ with $|A|<n$, there exists a unique smallest $W$-translate containing $A$.
%Consider a set $A \subseteq \mathbb{R}^2$ with $|A|=n$, suppose $A = \{a_1,\dots,a_n\}$.
Suppose $A = \{a_1,\dots,a_n\}$.
Let $\tilde{W} = \{(x,y):(-x,-y) \in W\}$, which is a wedge obtained by rotating $W$ around the origin with angle $\pi$.
Clearly, $a_i \in W_p$ iff $p \in \tilde{W}_{a_i}$.
Therefore, $A \subseteq W_p$ iff $p \in \bigcap_{i=1}^n \tilde{W}_{a_i}$.
Note that $\bigcap_{i=1}^n \tilde{W}_{a_i} = \tilde{W}_q$ for some $q \in \mathbb{R}^2$ as the intersection of finitely many $\tilde{W}$-translates is a $\tilde{W}$-translate.
We claim that $W_q$ is the smallest $W$-translate containing $A$.
It suffices to show that \textbf{(i)} $A \subseteq W_q$ and \textbf{(ii)} for any $q' \in \mathbb{R}^2$ with $A \subseteq W_{q'}$, $W_q \subseteq W_{q'}$.
We have $A \subseteq W_q$ since $q \in \bigcap_{i=1}^n \tilde{W}_{a_i} = \tilde{W}_q$.
Let $q' \in \mathbb{R}^2$ such that $A \subseteq W_{q'}$.
Then $q' \in \bigcap_{i=1}^n \tilde{W}_{a_i} = \tilde{W}_q$, which implies $q \in W_{q'}$ and $W_q \subseteq W_{q'}$.

Next, we prove the criterion given in the lemma.
Let $p \in \mathbb{R}^2$ be a point.
We denote by $r_p$ and $r_p'$ the $r$-branch and $r'$-branch of $W_p$, respectively.
To see ``if'', suppose $A \subseteq W_p$, $A \cap r_p \neq \emptyset$, and $A \cap r_p' \neq \emptyset$.
We claim that $W_p$ is the smallest $W$-translate containing $A$.
Let $W_{p'}$ be the smallest $W$-translate containing $A$.
Then we have $W_{p'} \in W_p$ and $p' \in W_p$.
If $p' \in W_p \backslash \{p\}$, then either $W_{p'} \cap r_p = \emptyset$ or $W_{p'} \cap r_p' = \emptyset$.
But $A \cap r_p \neq \emptyset$ and $A \cap r_p' \neq \emptyset$, which implies $A \nsubseteq W_{p'}$, contradicting the assumption that $W_{p'}$ contains $A$.
Thus, $p' = p$ and $W_p$ is the smallest $W$-translate containing $A$.
To see ``only if'', suppose $W_p$ is the smallest $W$-translate containing $A$.
Clearly, $A \subseteq W_p$.
It suffices to show that $A \cap r_p \neq \emptyset$ and $A \cap r_p' \neq \emptyset$.
Assume $A \cap r_p = \emptyset$.
Then one can always find a point $q \in r_p' \backslash \{p\}$ sufficiently close to $p$ such that $A \subseteq W_q$, contradicting the fact that $W_p$ is the smallest $W$-translate containing $A$.
Therefore, $A \cap r_p \neq \emptyset$, and for the same reason $A \cap r_p' \neq \emptyset$.

\subsection{Proof of Lemma~\ref{lem-associate}}
Recall that $r$ and $r'$ are the two branches of $C$.
Let $H$ be the downward-open halfplane bounded by $r$ and $H'$ be the leftward-open halfplane bounded by $r'$.
Then $C = H \cup H'$ and $C_p = H_p \cup H_p'$ for all $p \in \mathbb{R}^2$.

Let $q \in \mathbb{R}^2$ such that $\phi = (a,b)$ is the closest-pair in $S \cap C_q$.
%Recall that $C_q = H_q \cup H'_q$.
Note that $a$ and $b$ cannot be contained in $H_q$ simultaneously.
Indeed, if $a,b \in H_q$, then $\phi$ is the closest-pair in $S \cap H_q$ (as $S \cap H_q \subseteq S \cap C_q$) and thus $\phi \in \varPhi(S,\mathcal{H})$, contradicting the fact that $\phi \in \varPhi^*$.
For the same reason, $a$ and $b$ cannot be contained in $H'_q$ simultaneously.
So assume $a \in H_q \backslash H'_q$ and $b \in H'_q \backslash H_q$.
Draw a line $s$ parallel to $r$ (i.e., horizontal) through $a$, and draw a line $s'$ parallel to $r'$ through $b$.
Then $s$ and $s'$ intersect at a point $p \in H_q \cap H'_q$.
We claim that $C_p$ is the associated $C$-translate of $\phi$ and satisfies the conditions given in the lemma.
First, both $p$ and $a$ lie on $s$, and $p$ is to the left of $a$ (as $p \in H'_q$ and $a \notin H'_q$).
This implies that $a$ lies on the $r$-branch of $C_p$.
Using the same argument, we deduce that $b$ is on the $r'$-branch of $C_p$.
Therefore, $a,b \in C_p$ (and also $a,b \in W_p$).
Furthermore, $C_p \subseteq C_q$ because $H_p \subseteq H_q$ and $H'_p \subseteq H'_q$.
It follows that $\phi$ is the closest-pair in $S \cap C_p$.
To see that $C_p$ is the associated $C$-translate of $\phi$, we only need to show that $W_p$ is the smallest $W$-translate containing $a$ and $b$.
Since $a$ and $b$ lie on the two branches of $W_p$ respectively, by Lemma~\ref{lem-smallest}, $W_p$ is the smallest $W$-translate containing $a$ and $b$.
%Define $s$ and $s'$ as the lines obtained by extending the $r$-branch and $r'$-branch of $C_q$, respectively.
%By our assumption $r = \{(t,0): t \geq 0\}$ and $r' = \{(\alpha t,t): t \geq 0\}$, we can write $s: y = c$ and $s': x-\alpha y = c'$ for some $c,c' \in \mathbb{R}$.
%Let $H$ (resp., $H'$) be the halfplane $y \leq c$ (resp., $x - \alpha y \leq c'$), then $C_q = H \cup H'$.
\begin{figure}[h]
    \centering
    \includegraphics[height=2.8cm]{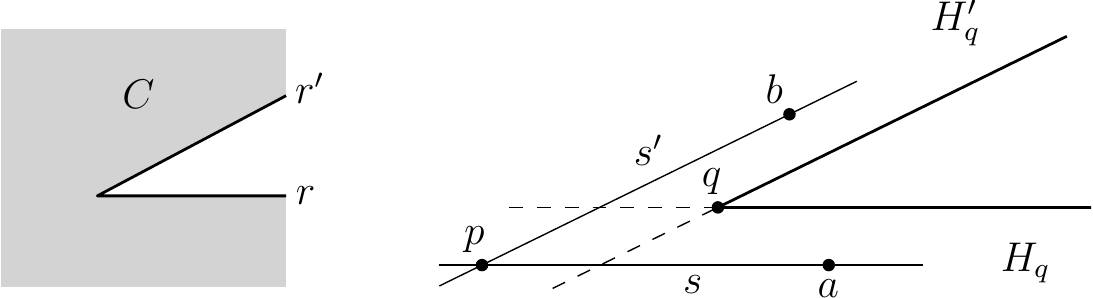}
    \caption{Illustrating the proof of Lemma~\ref{lem-associate}}
    \label{fig-prfass}
\end{figure}

\subsection{Proof of Lemma~\ref{lem-bipartite}}
Let $H$ be the downward-open halfplane bounded by $r$ and $H'$ be the leftward-open halfplane bounded by $r'$.
Then $C = H \cup H'$ and $C_p = H_p \cup H_p'$ for all $p \in \mathbb{R}^2$.

\begin{figure}[h]
    \centering
    \includegraphics[height=3cm]{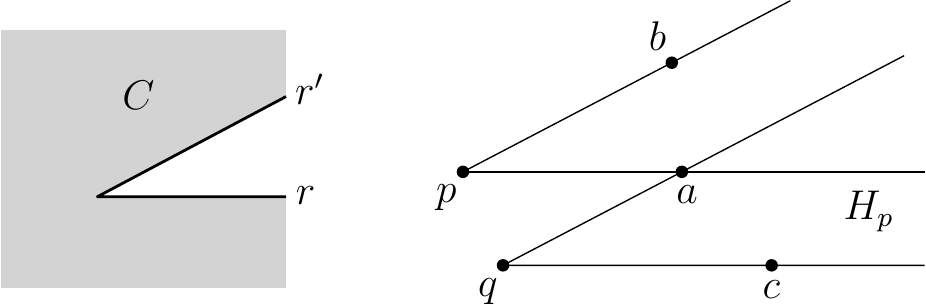}
    \caption{Illustrating the proof of Lemma~\ref{lem-bipartite}}
    \label{fig-prfbipartite}
\end{figure}

Assume $R \cap R' \neq \emptyset$ and $a \in R \cap R'$.
%Without loss of generality, assume $b$ is the origin of $\mathbb{R}^2$.
Since $a \in R$, there exists $\phi = (a,b) \in \varPhi^*$ such that $a$ (resp., $b$) lies on the $r$-branch (resp., $r'$-branch) of $C_p$, where $C_p$ is the associated $C$-translate of $\phi$.
Also, since $a \in R'$, there exists $\psi = (c,a) \in \varPhi^*$ such that $c$ (resp., $a$) lies on the $r$-branch (resp., $r'$-branch) of $C_q$, where $C_q$ is the associated $C$-translate of $\psi$.
We claim that $a,b,c \in C_p$ and $a,b,c \in C_q$.
It actually suffices to show $a,b,c \in C_p$.
Since $C_p$ is the associated $C$-translate of $\phi$, we have $a,b \in C_p$ by Lemma~\ref{lem-associate}.
To see $c \in C_p$, recall our assumption that $r = \{(t,0): t \geq 0\}$ and $r' = \{(\alpha t,t): t \geq 0\}$.
Because $a$ lies on the $r'$-branch of $C_q$ while $c$ lies on the $r$-branch, $a$ must be above $c$ (i.e., the $y$-coordinate of $a$ is greater than or equal to that of $c$).
See Figure~\ref{fig-prfbipartite}.
Therefore, $c \in H_p \subseteq C_p$ (as $H_p$ is the downward-open halfplane bounded by the horizontal line through $a$) and $a,b,c \in C_p$.
Using the same argument, we can deduce $a,b,c \in C_q$.
By Lemma~\ref{lem-associate}, $\phi$ is the closest-pair in $S \cap C_p$ and thus the closest-pair in $\{a,b,c\}$.
On the other hand, however, we also have that $\psi$ is the closest-pair in $\{a,b,c\}$ (as it is the closest-pair in $S \cap C_q$), which results in a contradiction.
As such, $R \cap R' = \emptyset$.
It follows directly that $G = (S,\varPhi^*)$ is a bipartite graph, since every pair (i.e., edge) in $\varPhi^*$ has exactly one point in $R$ and one point in $R'$.

\subsection{Proof of Lemma~\ref{lem-level}}
The proof of this statement follows from the probabilistic arguments of Clarkson and Shor~\cite{clarkson1989applications}, but we still present it here for the sake of completeness. 
The presentation  follows from the textbook of Har-Peled~\cite{har2011geometric}.
Pick a random sample $\mathcal{R} \subseteq \mathcal{S}$ where each object is picked with probability $1/k$. 
Therefore, $\mathbb{E}[|\mathcal{R}|]=n/k$.
We use $U$ to denote the union of the ranges in $\mathcal{R}$, and $r$ be the number of the vertices of $U$ which is a random variable.
Since $\mathcal{R}$ is a collection of pseudo-discs, the complexity of $U$ is $O(|\mathcal{R}|)$ \cite{agarwal2007shape_union} and hence $\mathbb{E}[r] = O(\mathbb{E}[|\mathcal{R}|]) = O(n/k)$.

Let $V_{\leq k}$ be the set of the vertices of the $\leq k$-level $\mathcal{A}_{\leq k}$.
For a vertex $v\in V_{\leq k}$, let $\delta_v$ be a random variable which is $1$ if $v$ is a vertex of $U$ and $0$ otherwise.
Note that $\delta_v$ is $1$ iff \textbf{(a)} the two ranges in $\mathcal{S}$ defining $v$ are both sampled in $\mathcal{R}$, and \textbf{(b)} none of the other ranges in $\mathcal{S}$ that contain $v$ is sampled in $\mathcal{R}$.
As such,
\begin{equation*}
    \mathbb{E}[\delta_v] = \Pr[\delta_v=1] \geq \left(1-\frac{1}{k} \right)^k\left(\frac{1}{k}\right)^2\geq e^{-2}\frac{1}{k^2}=\frac{1}{(ek)^2},
\end{equation*}
since $1-x \geq e^{-2x}$ for $x\in (0,1/2]$.
%\[ {\bf Pr}[X_v=1] \geq \left(1-\frac{1}{k} \right)^k\left(\frac{1}{k}\right)^2\geq e^{-2}\frac{1}{k^2}=\frac{1}{e^2k^2}, \text{ since $1-x \geq e^{-2x}$, if $x\in (0,1/2]$}.\]
Now the goal is to find an upper bound and a lower bound for the quantity $\sum_{v\in V_{\leq k}} \mathbb{E}[\delta_v]$.
For the lower bound, we have $\sum_{v \in V_{\leq k}} \mathbb{E}[\delta_v] \geq |V_{\leq k}|/(ek)^2$.
%\[{\bf E}\left[ \sum_{v\in V_{\leq k}({\cal S})}X_v \right] =\sum_{v\in V_{\leq k}({\cal S})} {\bf E}[X_v] \geq \frac{|V_{\leq k}({\cal S})|}{e^2k^2}\]
To establish an upper bound, we notice that $\sum_{v \in V_{\leq k}} \delta_v \leq r$.
Therefore,
\begin{equation*}
    \sum_{v \in V_{\leq k}} \mathbb{E}[\delta_v] = \mathbb{E} \left[\sum_{v\in V_{\leq k}} \delta_v \right] \leq \mathbb{E}[r] = O(n/k).
\end{equation*}
Combining the upper bound and the lower bound, we have $|V_{\leq k}|/(ek)^2 = O(n/k)$ and thus $|V_{\leq k}| = O(kn)$.
Therefore, the complexity of $\mathcal{A}_{\leq k}$ is $O(kn)$.
Since $k$ is a constant, we further conclude that the complexity of $\mathcal{A}_{\leq k}$ is $O(n)$.
 
\subsection{Proof of Lemma~\ref{lem-2shapes}}
\begin{figure}[htbp]
    \begin{center}
        \includegraphics[height=4.5cm]{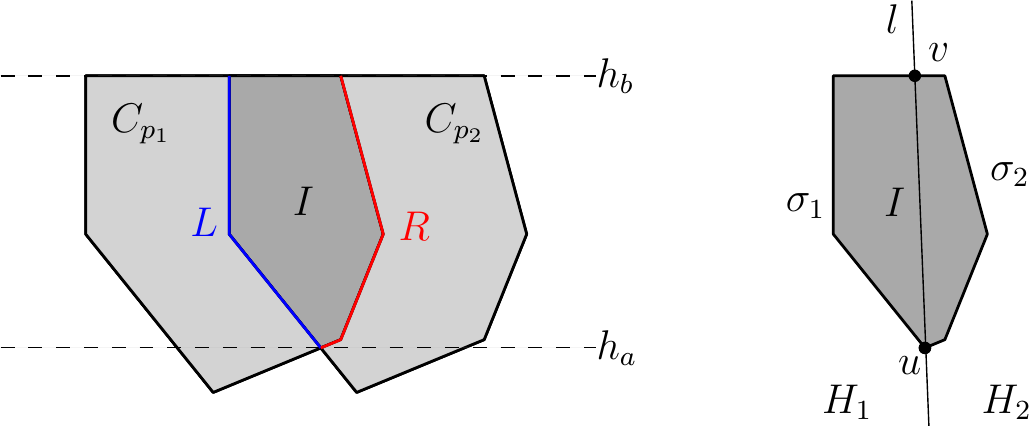}
    \end{center}
    \caption{Illustrating the proof of Lemma~\ref{lem-2shapes}}
    \label{fig-prf2}
\end{figure}

\ifdefined\SODA
\vspace{-0.15cm}
\fi
Without loss of generality, assume that $p_1 = (1,0)$ and $p_2 = (0,0)$.
Under this assumption, $C_{p_2}$ is obtained by ``moving'' $C_{p_1}$ leftward by distance $1$, and we have $\text{len}_{C_{p_1}}(h) = \text{len}_{C_{p_2}}(h) = \text{len}_C(h)$ for any horizontal line $h$.
Let $h_t$ be the horizontal line $x = t$ for $t \in \mathbb{R}$.
Then it is easy to see that $I \cap h_t \neq \emptyset$ iff $\text{len}_C(h_t) \geq 1$.
Let $T = \{t:I \cap h_t \neq \emptyset\}$, $a = \inf T$, $b = \sup T$.
Clearly, $T=[a,b]$.
We denote by $L$ (resp., $R$) the set of the left (resp., right) endpoints of the segments $I \cap h_t$ for $t \in T$; see the left figure of Figure~\ref{fig-prf2}.
Since $I$ is a convex body, $L$ and $R$ must be connected (intuitively, $L$ is the left ``boundary'' of $I$ while $R$ is the right ``boundary'').
Define $L' = L \cup (I \cap h_a) \cup (I \cap h_b)$ and $R' = R \cup (I \cap h_a) \cup (I \cap h_b)$.
One can also easily verify the connectedness of $L'$ and $R'$.

To prove that $C_{p_1}$ and $C_{p_2}$ plainly intersect, suppose $I \neq \emptyset$.
We claim that $\partial C_{p_1} \cap C_{p_2} = L'$ and $\partial C_{p_2} \cap C_{p_1} = R'$.
It suffices to show $\partial C_{p_1} \cap C_{p_2} = L'$.
First, we observe that $L \subseteq \partial C_{p_1}$.
Indeed, the left endpoint of the segment $I \cap h_t$ for any $t \in T$ must be the left endpoint of $C_{p_1} \cap h_t$.
Then we show that $I \cap h_a \subseteq \partial C_{p_1}$.
To this end, we only need to consider the case that $\text{len}_I(h_a) > 0$ (otherwise $I \cap h_a$ contains a single point and $I \cap h_a \subseteq L \subseteq \partial C_{p_1}$).
Note that $\text{len}_I(h_t) = \max\{\text{len}_C(h_t) - 1, 0\}$ for all $t \in \mathbb{R}$.
Therefore, $\text{len}_C(h_a) > 1$.
But $\text{len}_C(h_t) < 1$ for all $t < a$, as $I \cap h_t = \emptyset$ for all $t < a$.
This situation happens only when $C$ (as well as $C_{p_1}$ and $C_{p_2}$) is entirely on the top side of $h_a$.
It follows that $I \cap h_a \subseteq C_{p_1} \cap h_a \subseteq \partial C_{p_1}$.
Using the same argument, we can show $I \cap h_b \subseteq \partial C_{p_1}$.
Thus, $L' \subseteq \partial C_{p_1}$.
In fact, $L' \subseteq \partial C_{p_1} \cap C_{p_2}$ because $L' \subseteq I \subseteq C_{p_2}$.
The remaining task is to show that $\partial C_{p_1} \cap C_{p_2} \subseteq L'$.
To this end, we first observe that $\partial C_{p_1} \cap C_{p_2} \subseteq \partial I$.
Indeed, any point in $\partial C_{p_1} \cap C_{p_2}$ must be in $I$, and must be not in $I^\circ$ (for it is in $\partial C_{p_1}$).
With this observation, it is sufficient to show that $\partial C_{p_1} \cap C_{p_2} \cap (\partial I \backslash L') = \emptyset$.
Take a point $r \in \partial I \backslash L'$.
Then $r$ must be the right endpoint of a segment $I \cap h_t$ for some $t \in (a,b)$.
Furthermore, since $r$ is not the left endpoint of $I \cap h_t$ (otherwise $r \in L'$), we have $\text{len}_I(h_t) > 0$, which implies $\text{len}_C(h_t) > 1$.
In this situation, it is easy to see that $r \in C_{p_1}^\circ$.
It follows that $\partial C_{p_1} \cap (\partial I \backslash L') = \emptyset$ and thus $\partial C_{p_1} \cap C_{p_2} \cap (\partial I \backslash L') = \emptyset$.
Therefore, $\partial C_{p_1} \cap C_{p_2} \subseteq L'$ and $\partial C_{p_1} \cap C_{p_2} = L'$.
For the same reason, we have that $\partial C_{p_2} \cap C_{p_1} = R'$.
Now both $\partial C_{p_1} \cap C_{p_2}$ and $\partial C_{p_2} \cap C_{p_1}$ are connected, so $C_{p_1}$ and $C_{p_2}$ plainly intersect.

The rest of the proof is dedicated to find the points $u,v$ satisfying the desired properties.
Suppose $I^\circ \neq \emptyset$.
We define $u$ (resp., $v$) as an arbitrary point in $I \cap h_a$ (resp., $I \cap h_b$).
Let $l$ be the line through $u,v$.
The property (3) is clearly satisfied. 
Since $I^\circ \neq \emptyset$, we have $a \neq b$ and thus (2) is satisfied.
To see (4), consider the two arcs $\sigma_1,\sigma_2$ in $\partial I$ connecting $u,v$; see the right figure of Figure~\ref{fig-prf2}.
Clearly, $L$ is contained in one arc (say $L \subseteq \sigma_1$) while $R$ is contained in the other one (say $R \subseteq \sigma_2$).
Then $\sigma_1 \subseteq L' \subseteq \partial C_{p_1}$ and $\sigma_2 \subseteq R' \subseteq \partial C_{p_2}$.
Finally, we show that (1) is also satisfied.
Let $H_1$ (resp., $H_2$) be the left (resp., right) side of $l$ (this makes sense as $l$ is not horizontal).
We have $\sigma_i \subseteq H_i$ for $i \in \{1,2\}$.
We only need to show that $H_1$ is $I$-vanishing, for the roles of $H_1$ and $H_2$ are symmetric.
If $\sigma_1 \subseteq l$, then $I \subseteq H_2$ and we are done.
If $\sigma_1 \nsubseteq l$, then we must have $I \cap H_1 = C_{p_1} \cap H_1$, because both sides of the equation are equal to $\mathcal{CH}(\sigma_1)$ in this case.
As such, $\text{len}_{C_{p_1}}(l) = \text{len}_I(l) \leq \text{len}_{C_{p_2}}(l)$.
Since $I \cap H_1 = C_{p_1} \cap H_1$, it suffices to show that $H_1$ is $C_{p_1}$-vanishing.
For any point $z \in \mathbb{R}^2$, we define $d(z) = -\text{dist}(z,l)$ if $z \in H_1$ and $d(z) = \text{dist}(z,l)$ otherwise, where $\text{dist}(z,l)$ is the distance from $z$ to $l$.
Observing the locations of $p_1,p_2$, we have that $d(p_1) > d(p_2)$.
Let $l'$ be the line parallel to $l$ such that $l' \subseteq H_2$ and $\text{dist}(l,l') = d(p_1)-d(p_2)$, where $\text{dist}(l,l')$ is the distance between the parallel lines $l$ and $l'$.
Now it is easy to verify that $\text{len}_{C_{p_1}}(l') = \text{len}_{C_{p_2}}(l)$.
Using the previous observation that $\text{len}_{C_{p_1}}(l) \leq \text{len}_{C_{p_2}}(l)$, we have $\text{len}_{C_{p_1}}(l) \leq \text{len}_{C_{p_1}}(l')$.
However, this implies that $H_2$ is not strictly $C_{p_1}$-vanishing, and thus $H_1$ must be $C_{p_1}$-vanishing.

\ifdefined\SODA
\vspace{-0.25cm}
\fi
\subsection{Proof of Corollary~\ref{cor-halfcontain}}
\begin{figure}[htbp]
    \begin{center}
        \includegraphics[height=4.5cm]{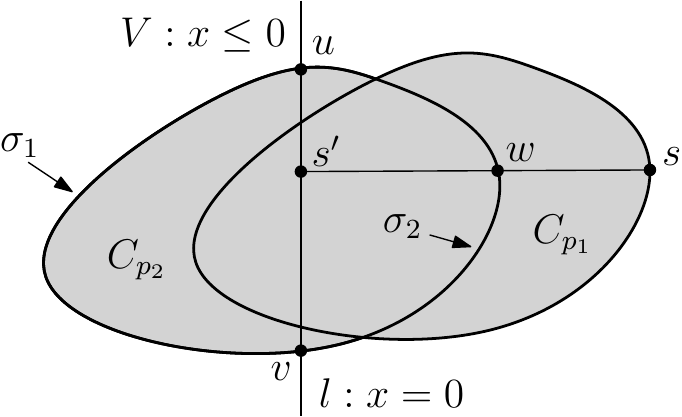}
    \end{center}
    \caption{Illustrating the proof of Corollary~\ref{cor-halfcontain}}
    \label{fig-prf3}
\end{figure}

\ifdefined\SODA
\vspace{-0.15cm}
\fi
Let $I = C_{p_1} \cap C_{p_2}$.
First, we prove the statement (1).
Due to the convexity of $C_{p_2}$, it suffices to show that $\partial C_{p_1} \cap V \subseteq C_{p_2}$.
The statement is trivial if $I = \emptyset$, so suppose $I \neq \emptyset$. 
Then by Lemma~\ref{lem-2shapes}, $C_{p_1}$ and $C_{p_2}$ plainly intersect, hence $\partial C_{p_1} \cap C_{p_2}$ is connected.
It follows that either $\partial C_{p_1} \cap V \subseteq C_{p_2}$ or $\partial C_{p_1} \backslash V \subseteq C_{p_2}$, because $C_{p_1} \cap l \subseteq C_{p_2}$.
We show that $\partial C_{p_1} \backslash V \nsubseteq C_{p_2}$.
Without loss of generality, assume $l$ is the line $x=0$ and $V$ is the side $x \leq 0$.
Let $s$ be a rightmost point in $C_{p_1}$, i.e., a point with a maximum abscissa; see Figure~\ref{fig-prf3}.
Then $s \in \partial C_{p_1}$.
Since $\text{len}_{C_{p_1}}(l) < \text{len}_{C_{p_2}}(l)$ and $V$ is a $C_{p_1}$-vanishing side, we have $p_1.x>p_2.x$.
This implies that $s$ has a greater abscissa than any point in $C_{p_2}$.
In particular, $s \notin C_{p_2}$.
%On the other hand, $s \in \partial C_{p_1}$ for $s$ is a rightmost point in $C_{p_1}$.
Furthermore, we have $s.x>0$ (i.e., $s \notin V$), because $C_{p_2} \cap l \neq \emptyset$ and the abscissas of the points in $C_{p_2} \cap l$ are 0.
Thus, $s \in \partial C_{p_1} \backslash V$.
As $s \notin C_{p_2}$ and $s \in \partial C_{p_1} \backslash V$, we deduce that $\partial C_{p_1} \backslash V \nsubseteq C_{p_2}$, implying $\partial C_{p_1} \cap V \subseteq C_{p_2}$.
Next, we prove the statement (2).
The statement (1) already implies $C_{p_1} \cap (V \backslash l) \subseteq C_{p_2}$, thus it suffices to show that $\partial C_{p_2} \cap C_{p_1} \cap (V \backslash l) = \emptyset$.
If $C_{p_2} \cap (V \backslash l) = \emptyset$, we are done.
So suppose $C_{p_2} \cap (V \backslash l) \neq \emptyset$, then $\partial C_{p_2} \cap (V \backslash l) \neq \emptyset$.
Let $u,v$ be the two endpoints of $C_{p_2} \cap l$.
Take a point $r \in \partial C_{p_2} \cap (V \backslash l)$.
There are two arcs in $\partial C_{p_2}$ connecting $u,v$, say $\sigma_1$ and $\sigma_2$.
Assume $\sigma_1 \subseteq V$ and $\sigma_2$ is on the other side of $l$ than $V$.
Then $r \in \sigma_1$, and $\sigma_1 \cap l = \{u,v\}$.
We claim that $\sigma_2 \cap C_{p_1} \neq \emptyset$.
We assume as before that $l$ is the line $x=0$ and $V$ is the side $x \leq 0$.
Let $s$ be a rightmost point in $C_{p_1}$.
As argued in the proof of the statement (1), $s.x>0$ and $s \notin C_{p_2}$.
Now take another point $s' \in C_{p_1} \cap l$.
Since $s \notin C_{p_2}$ and $s' \in C_{p_2}$, there must exist a point $w$ on the segment $[s,s']$ such that $w \in \partial C_{p_2}$; see Figure~\ref{fig-prf3}.
Note that $w \in \partial C_{p_2} \cap C_{p_1}$, as $s,s' \in C_{p_1}$.
Furthermore, as $s.x \geq 0$ and $s'.x \geq 0$, we have $w.x \geq 0$, i.e., $w \notin V \backslash l$.
%Furthermore, $w \notin \{u,v\}$, since $C_{p_1} \cap l \subseteq (C_{p_2} \cap l) \backslash \{u,v\}$.
So we deduce that $w \in \sigma_2 \cap C_{p_1}$, i.e., $\sigma_2 \cap C_{p_1} \neq \emptyset$.
This implies $\partial C_{p_2} \cap C_{p_1} \subseteq \sigma_2$, because $u,v \notin C_{p_1}$ and $\partial C_{p_2} \cap C_{p_1}$ is connected by Lemma~\ref{lem-2shapes}.
Therefore, $\partial C_{p_2} \cap C_{p_1} \cap (V \backslash l) \subseteq \sigma_2 \cap (V \backslash l) = \emptyset$, as desired.

\ifdefined\SODA
\vspace{-0.25cm}
\fi
\subsection{Proof of Lemma~\ref{lem-contain}}
\begin{figure}[htbp]
    \begin{center}
        \includegraphics[height=4cm]{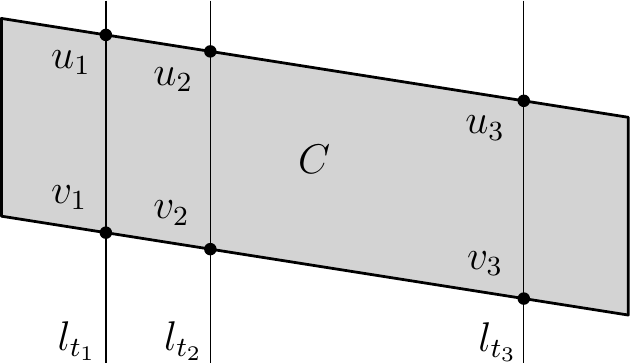}
    \end{center}
    \caption{Illustrating the notations in the proof of Fact~\ref{fact-degen}}
    \label{fig-prf4}
\end{figure}

\ifdefined\SODA
\vspace{-0.15cm}
\fi
In order to prove Lemma~\ref{lem-contain}, we first establish the following basic fact.
This fact will also be used in the proof of Lemma~\ref{lem-exclude} and Lemma~\ref{lem-3shapes}.
\ifdefined\SODA
\vspace{-0.1cm}
\fi
\begin{fact} \label{fact-degen}
	Let $C,p_1,p_2$ be as in Lemma~\ref{lem-2shapes} such that $I^\circ \neq \emptyset$ for $I = C_{p_1} \cap C_{p_2}$.
	Also, let $u,v,l$ be the points and line satisfying the conditions in Lemma~\ref{lem-2shapes}.
	If $I \cap l = C_p \cap l$ for some $p \in \mathbb{R}^2$, then $p,p_1,p_2$ are collinear.
\end{fact}
\ifdefined\SODA
\vspace{-0.1cm}
\fi
\textit{Proof.}
Without loss of generality, we may assume $u = (0,1)$ and $v = (0,0)$.
Then $l$ is the line $x = 0$.
By (4) of Lemma~\ref{lem-2shapes}, the two arcs in $\partial I$ connecting $u,v$ are contained in $\partial C_{p_1}$ and $\partial C_{p_2}$ respectively (we denote by $\sigma_i$ the arc contained in $C_{p_i}$ for $i \in \{1,2\}$).
Let $H_1,H_2$ be the two sides of $l$ such that $\sigma_i \subseteq H_i$ for $i \in \{1,2\}$.
Since $I^\circ \neq \emptyset$, at least one of $\sigma_1,\sigma_2$ is not contained in $l$ (say $\sigma_1 \neq l$).
Then $I \cap H_1 = C_{p_1} \cap H_1$, because both sides of the equation are equal to $\mathcal{CH}(\sigma_1)$ when $\sigma_1 \neq l$.
In particular, $C_{p_1} \cap l = I \cap l = C_p \cap l$.
We consider two cases, $C_{p_2} \cap l \neq I \cap l$ and $C_{p_2} \cap l = I \cap l$.
If $C_{p_2} \cap l \neq I \cap l$, then $I \cap l \subsetneq C_{p_2} \cap l$.
In this case, we must have $\sigma_2 \subseteq l$ (otherwise $I \cap H_2 = C_{p_2} \cap H_2$ and $I \cap l = C_{p_2} \cap l$), i.e., $\sigma_2$ is the segment $[u,v]$.
Since $\sigma_2$ is a portion of $\partial C_{p_2}$, $C_{p_2}$ is entirely on one side of $l$ (which must be $H_1$).
We claim that there exists a unique $t \in \mathbb{R}$ such that $\text{len}_C(l_t) = 1$, where $l_t$ denotes the line $x = t$.
Assume now $H_1$ is the side $x \leq 0$.
The existence is clear, because $\text{len}_{C_{p_1}}(l) = \text{len}_I(l) = \text{dist}(u,v) = 1$.
To see the uniqueness, assume there exist $t_1,t_2 \in \mathbb{R}$ such that $t_1<t_2$ and $\text{len}_C(l_{t_1}) = \text{len}_C(l_{t_2}) = 1$.
Let $t^* = \sup \{t \in \mathbb{R}: C \cap l_t \neq \emptyset\}$, then $t^* \geq t_2>t_1$.
Note that $\text{len}_C(l_{t^*}) = \text{len}_{C_{p_2}}(l)$.
Since $I \cap l \subsetneq C_{p_2} \cap l$, we have $\text{len}_{C_{p_2}}(l) > \text{len}_{C_{p_1}}(l) = 1$.
It follows that $t^* \neq t_2$ and thus $t^* > t_2$.
Now $\text{len}_C(l_{t^*}) > 1 = \text{len}_C(l_{t_2})$, so the side of $l_{t_2}$ containing $l_{t^*}$ is not $C$-vanishing, which implies that the other side of $l_{t_2}$ (i.e., the side containing $l_{t_1}$) is strictly $C$-vanishing.
But this contradict the fact that $\text{len}_C(l_{t_1})=1$.
Therefore, there exists a unique $t \in \mathbb{R}$ such that $\text{len}_C(l_t) = 1$.
%With this in hand, we show that $p_1 = p$.
Using this observation and the fact that $\text{len}_{C_{p_1}}(l) = \text{len}_{C_p}(l) = 1$, we further deduce that $p_1.x = p.x$.
This implies $p_1.y = p.y$, since $C_{p_1} \cap l = C_p \cap l$.
Therefore, $p_1 = p$ and $p,p_1,p_2$ are collinear.
Next, we consider the case that $C_{p_2} \cap l = I \cap l$.
In this case, $C_{p_1} \cap l = C_{p_2} \cap l = C_p \cap l = [s,t]$ and $\text{len}_{C_{p_1}}(l) = \text{len}_{C_{p_2}}(l) = \text{len}_{C_p}(l) = 1$.
As argued before, if any two points in $\{p,p_1,p_2\}$ have the same abscissa, then they must have the same ordinate and must coincide, which implies $p,p_1,p_2$ are collinear.
So suppose $p,p_1,p_2$ have distinct abscissas $x_0,x_1,x_2$, respectively.
Let $t_i = -x_i$ for $i \in \{0,1,2\}$.
Clearly, $\text{len}_{C_{p_i}}(l) = \text{len}_C(l_{t_i})$ for $i \in \{0,1,2\}$.
We denote by $u_i$ (resp., $v_i$) the top (resp., bottom) endpoints of the segment $C \cap l_{t_i}$ for $i \in \{0,1,2\}$; see Figure~\ref{fig-prf4}.
Then we have the equations $u_i + p_i = u$ and $v_i + p_i = v$ for $i \in \{0,1,2\}$, where $p_0 = p$.
To show $p,p_1,p_2$ are collinear, it suffices to show that $u_0,u_1,u_2$ (or equivalently, $v_0,v_1,v_2$) are collinear.
Assume $t_0 < t_1 < t_2$.
If $u_1$ is above the segment $[u_0,u_2]$, then $v_1$ is in the interior of the triangle $\triangle v_0 u_1 v_2$ and hence in $C^\circ$, contradicting the fact that $v_1 \in \partial C$.
Similarly, if $u_1$ is below $[u_0,u_2]$, then $u_1$ is in the interior of triangle $\triangle u_0 v_1 u_2$ and hence in $C^\circ$, contradicting the fact that $u_1 \in \partial C$.
Therefore, we have $u_1 \in [u_0,u_2]$, which implies $u_0,u_1,u_2$ are collinear.
Note that the assumption $t_0 < t_1 < t_2$ is not necessary: the same argument applies no matter what the order of $t_0,t_1,t_2$ is.
\hfill $\Box$
\smallskip

\noindent
Now we prove Lemma~\ref{lem-contain}.
Suppose $I \cap l \subseteq C_p$ (then $I \cap l \subseteq C_p \cap l$).
Since $p,p_1,p_2$ are not collinear, by Fact~\ref{fact-degen} (see above) we have $I \cap l \subsetneq C_p \cap l$.
According to (1) and (4) of Lemma~\ref{lem-2shapes}, $l$ has two $I$-vanishing sides (say $H_1$ and $H_2$), and the two arcs in $\partial I$ connecting $u,v$ are contained in $\partial C_{p_1}$ and $\partial C_{p_2}$ respectively (we denote by $\sigma_i$ the arc contained in $C_{p_i}$ for $i \in \{1,2\}$).
%Note that $\partial I = \sigma_1 \cup \sigma_2$.
Without loss of generality, assume $\sigma_i \subseteq H_i$ for $i \in \{1,2\}$.
To show $I \subseteq C_p$, it suffices to show $I \cap H_1 \subseteq C_p$, as $I \cap H_2 \subseteq C_p$ can be proved symmetrically.
If $\sigma_1 \subseteq l$, then $I \cap H_1 = I \cap l \subseteq C_p$.
If $\sigma_1 \nsubseteq l$, we must have $I \cap H_1 = C_{p_1} \cap H_1$, because both sides of the equation are equal to $\mathcal{CH}(\sigma_1)$ in this case.
In particular, $C_{p_1} \cap l = I \cap l \subsetneq C_p \cap l$.
By (1) of Corollary~\ref{cor-halfcontain}, this implies $C_{p_1} \cap H_1 \subseteq C_p$.
It follows that $I \cap H_1 \subseteq C_p$ and thus $I \subseteq C_p$.
Next, suppose $I \cap l \subseteq C_p^\circ$.
By the above argument, we already have $I \subseteq C_p$.
So it suffices to show $I \cap \partial C_p = \emptyset$.
Assume $I \cap \partial C_p \neq \emptyset$ and take a point $r \in I \cap \partial C_p$.
Since $u,v \in C_p^\circ$ and $r \in \partial C_p$, it is possible to choose a point $p'$ in a sufficiently small neighborhood around $p$ such that $u,v \in C_{p'}^\circ$ but $r \notin C_{p'}$.
However, by the previous proof, if $u,v \in C_{p'}$, then $I \subseteq C_{p'}$, contradicting the fact that $r \notin C_{p'}$.
Therefore, $I \cap \partial C_p = \emptyset$ and $I \subseteq C_p^\circ$.

\ifdefined\SODA
\vspace{-0.25cm}
\fi
\subsection{Proof of Lemma~\ref{lem-exclude}}
\begin{figure}[htbp]
    \begin{center}
        \includegraphics[height=4.5cm]{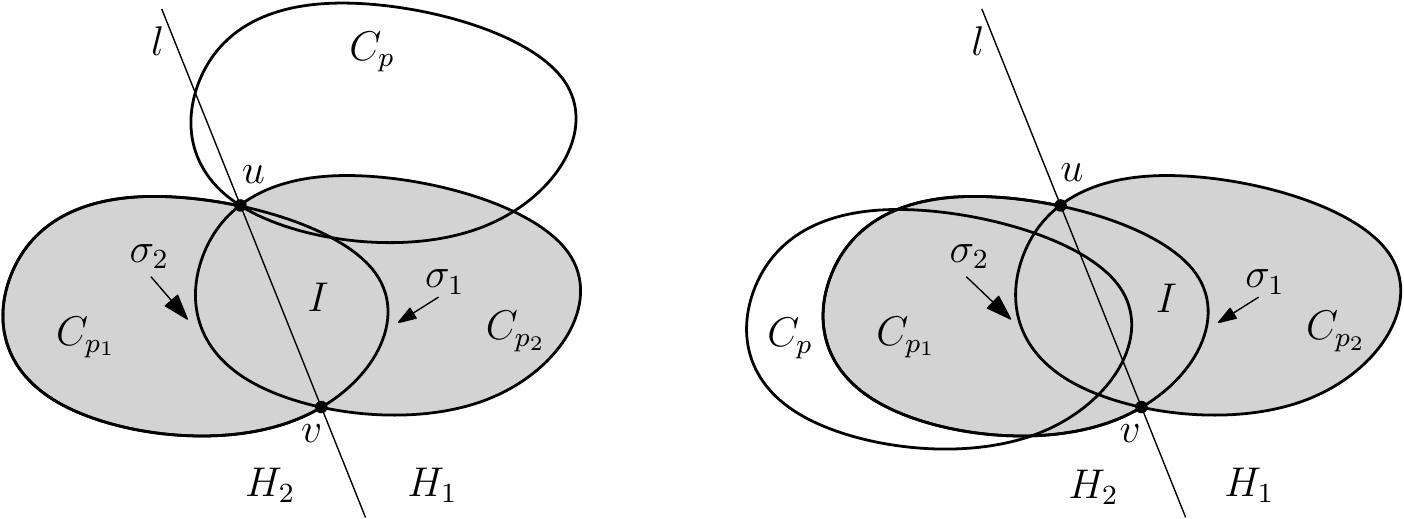}
    \end{center}
    \caption{Illustrating the the proof of Lemma~\ref{lem-exclude}}
    \label{fig-prf5}
\end{figure}

\ifdefined\SODA
\vspace{-0.15cm}
\fi
By (1) and (4) of Lemma~\ref{lem-2shapes}, $l$ has two $I$-vanishing sides (say $H_1$ and $H_2$), and the two arcs in $\partial I$ connecting $u,v$ are contained in $\partial C_{p_1}$ and $\partial C_{p_2}$ respectively (we denote by $\sigma_i$ the arc contained in $C_{p_i}$ for $i \in \{1,2\}$).
Without loss of generality, assume $\sigma_i \subseteq H_i$ for $i \in \{1,2\}$.
%(note that the two arcs must be contained in the two sides of $l$ respectively due to the convexity of $I$).
We first handle a special case in which $C_p \subseteq H_i$ for some $i \in \{1,2\}$, say $C_p \subseteq H_1$.
If $\sigma_1 \subseteq l$, then $I \subseteq H_2$ and thus $C_p \cap I \subseteq l$, contradicting the assumption that $I^\circ \cap C_p^\circ \neq \emptyset$.
If $\sigma_1 \nsubseteq l$, we must have $I \cap H_1 = C_{p_1} \cap H_1$, because both sides of the equation are equal to $\mathcal{CH}(\sigma_1)$ in this case.
Since $C_p \subseteq H_1$ by assumption, it follows that $C_p \cap I = C_p \cap I \cap H_1 = C_p \cap C_{p_1} \cap H_1 = C_p \cap C_{p_1}$.
In the rest of the proof, we may assume that $C_p \nsubseteq H_i$ for all $i \in \{1,2\}$.
%Let $s,t$ be the two endpoints of $C_p \cap l$.
Note that this assumption implies $(C_p \cap l) \backslash \{s,s'\} \subseteq C_p^\circ$, where $s,s'$ are the two endpoints of $C_p \cap l$.
%We consider the relation between the two segments $C_p \cap l$ and $I \cap l$.
We consider two cases separately: $C_p \cap l \nsubseteq I \cap l$ and $C_p \cap l \subseteq I \cap l$.
Suppose $C_p \cap l \nsubseteq I \cap l$.
Because $u,v \notin C_p^\circ$ and $(C_p \cap l) \backslash \{s,s'\} \subseteq C_p^\circ$, we have either $C_p \cap I \cap l = \emptyset$ or $C_p \cap I \cap l$ contains a single point (which must be $u$ or $v$); see the left figure of Figure~\ref{fig-prf5} for an illustration of the latter case.
In either of the two possibilities, $\text{len}_{C_p \cap I}(l) = 0$.
Since $(C_p \cap I)^\circ = I^\circ \cap C_p^\circ \neq \emptyset$, it follows that $C_p \cap I \subseteq H_i$ for some $i \in \{1,2\}$, say $C_p \cap I \subseteq H_1$.
Based on this, we claim that $C_p \cap I = C_p \cap C_{p_1}$.
If $\sigma_1 \subseteq l$, then $I \subseteq H_2$ and thus $C_p \cap I \subseteq H_1 \cap H_2 = l$, contradicting the assumption that $(C_p \cap I)^\circ \neq \emptyset$.
Otherwise, $\sigma_1 \nsubseteq l$ and thus $I \cap H_1 = \mathcal{CH}(\sigma_1) = C_{p_1} \cap H_1$.
Then we have $C_p \cap I = C_p \cap I \cap H_1 = C_p \cap C_{p_1} \cap H_1$.
Now it suffices to show that $C_p \cap C_{p_1} \cap H_1 = C_p \cap C_{p_1}$, i.e., $C_p \cap C_{p_1} \subseteq H_1$.
Since $C_p \cap I = C_p \cap C_{p_1} \cap H_1$, we have $(C_p \cap C_{p_1} \cap H_1)^\circ \neq \emptyset$.
But $\text{len}_{C_p \cap C_{p_1}}(l) = \text{len}_{C_p \cap I}(l) = 0$ (as $I \cap l = C_{p_1} \cap l$).
Therefore, $C_p \cap C_{p_1} \subseteq H_1$, as desired.
This completes the case that $C_p \cap l \nsubseteq I \cap l$.
Next, suppose $C_p \cap l \subseteq I \cap l$; see the right figure of Figure~\ref{fig-prf5}.
%Since $u,v \notin C_p^\circ$ and $(C_p \cap l) \backslash \{s,t\} \subseteq C_p^\circ$, there are three possibilities: $C_p \cap l \subseteq I \cap l$, $C_p \cap I \cap l = \emptyset$, $C_p \cap I \cap l = \{u\}$ (or $C_p \cap I \cap l = \{v\}$).
In this case, we must have $C_p \cap l \subsetneq I \cap l$ (otherwise, $p,p_1,p_2$ are collinear by Fact~\ref{fact-degen}, contradicting our assumption).
Without loss of generality, assume that $H_1$ is $C_p$-vanishing.
By (1) of Corollary~\ref{cor-halfcontain}, we have $C_p \cap H_1 \subseteq C_{p_i}$ for all $i \in \{1,2\}$, i.e., $C_p \cap H_1 \subseteq I$.
Therefore, $C_p \cap H_1 = C_p \cap I \cap H_1 \subseteq C_p \cap C_{p_2} \cap H_1 \subseteq C_p \cap H_1$, which implies $C_p \cap I \cap H_1 = C_p \cap C_{p_2} \cap H_1$.
On the other hand, we have $I \backslash H_1 = C_{p_2} \backslash H_1$, because $\partial I \backslash H_1 = \sigma_2 \backslash H_1 = \partial C_{p_2} \backslash H_1$.
It then follows that $C_p \cap (I \backslash H_1) = C_p \cap (C_{p_2} \backslash H_1)$.
By combining the observations that $C_p \cap I \cap H_1 = C_p \cap C_{p_2} \cap H_1$ and $C_p \cap (I \backslash H_1) = C_p \cap (C_{p_2} \backslash H_1)$, we can finally deduce $C_p \cap C_{p_2} = C_p \cap I$.

\subsection{Proof of Lemma~\ref{lem-tau}}
We first discuss the choice of the number $\tau$.
Since $C$ is smooth, for every $a \in \partial C$ there is a unique tangent line $\mathsf{tgt}(a)$ of $C$ through $a$.
Let $\mathbf{v}_a$ be the \textit{outward-pointing normal} at $a$, which is by definition a unit vector with initial point $a$ whose direction is perpendicular to $\mathsf{tgt}(a)$ and pointing to the exterior of $C$.
Define a mapping $f: \partial C \rightarrow \mathbb{S}^1$ as $f(a) = \mathbf{v}_a$, where $\mathbb{S}^1$ is the unit circle.
One can easily verify that $f$ is a continuous surjective map from $\partial C$ to $\mathbb{S}^1$.
For two points $\mathbf{u},\mathbf{u}' \in \mathbb{S}^1$, we denote by $\text{ang}(\mathbf{u},\mathbf{u}')$ the angle between $\mathbf{u}$ and $\mathbf{u}'$ (when regarded as unit vectors).
At each point $\mathbf{v} \in \mathbb{S}^1$, we take a sufficiently small open neighborhood $U_\mathbf{v} \subseteq \mathbb{S}^1$ such that $\text{ang}(\mathbf{u},\mathbf{u}') < \pi/2$ for any $\mathbf{u},\mathbf{u}' \in U_\mathbf{v}$.
Then $\mathcal{U} = \{U_\mathbf{v}: \mathbf{v} \in \mathbb{S}^1\}$ is an open cover of $\mathbb{S}^1$, and $f^{-1}(\mathcal{U}) = \{f^{-1}(U_\mathbf{v}): \mathbf{v} \in \mathbb{S}^1\}$ is an open cover of $\partial C$.
Now we regard $\partial C$ as a metric space equipped with the Euclidean metric.
Since $\partial C$ is compact, by Lebesgue's number lemma, there exists a number $\delta > 0$ such that any subset of $\partial C$ having diameter less than $\delta$ is contained in some member of the open cover $f^{-1}(\mathcal{U})$.
We then choose $\tau = \delta$.

Next, we verify that the number $\tau$ chosen above satisfies the desired property.
Let $l$ be a line such that $0 < \text{len}_C(l) < \tau$, and $H$ be a $C$-vanishing side of $l$.
We denote by $e_1,e_2$ the two intersection point of $\partial C$ and $l$, then $C \cap l = [e_1,e_2]$.
Let $r \in C \cap H$ be a point, and our goal is to show that $\text{dist}(r,e_1) < \text{len}_C(l)$ and $\text{dist}(r,e_2) < \text{len}_C(l)$.
Since $\text{dist}(e_1,e_2) = \text{len}_C(l) < \tau$, i.e., the diameter of the set $\{e_1,e_2\}$ is less than $\tau$, we have $e_1,e_2 \in f^{-1}(U_\mathbf{v})$ for some $\mathbf{v} \in \mathbb{S}^1$ (according to our choice of $\tau$).
It follows that $\mathbf{v}_{e_1}, \mathbf{v}_{e_2} \in U_\mathbf{v}$.
Recall that $\text{ang}(\mathbf{u},\mathbf{u}') < \pi/2$ for any $\mathbf{u},\mathbf{u}' \in U_\mathbf{v}$.
Therefore, $\text{ang}(\mathbf{v}_{e_1},\mathbf{v}_{e_2}) < \pi/2$.
Let $l_1 = \mathsf{tgt}(e_1)$ and $l_2 = \mathsf{tgt}(e_2)$ be the tangent lines of $C$ through $e_1$ and $e_2$ respectively.
Note that $l_1$ and $l_2$ are not parallel; indeed, if they are parallel, then $\text{ang}(\mathbf{v}_{e_1},\mathbf{v}_{e_2}) = \pi$.
Furthermore, the intersection point $o$ of $l_1$ and $l_2$ must lie in $H$ because $H$ is $C$-vanishing.
Consider the triangle $\triangle e_1 o e_2$.
Clearly, we have $C \cap H \subseteq \triangle e_1 o e_2$, and thus $r \in \triangle e_1 o e_2$.
Now $\angle e_1 r e_2 \geq \angle e_1 o e_2 = \pi - \text{ang}(\mathbf{v}_{e_1},\mathbf{v}_{e_2}) > \pi/2$.
Therefore, $\text{dist}(r,e_1) < \text{dist}(e_1,e_2) = \text{len}_C(l)$ and $\text{dist}(r,e_2) < \text{dist}(e_1,e_2) = \text{len}_C(l)$, which completes the proof.

%insert here
\subsection{Proof of Lemma~\ref{lem-3shapes}}
\ifdefined\SODA
\vspace{-0.15cm}
\fi
If $I^\circ \cap C_p^\circ = \emptyset$, then $\partial C_p \cap I = \partial I \cap C_p = C_p \cap I$, which is clearly connected.
So suppose $I^\circ \cap C_p^\circ \neq \emptyset$.
We need to show that $\partial C_p \cap I$ and $\partial I \cap C_p$ are connected.
Let $u,v,l$ be the points and line satisfying the conditions in Lemma~\ref{lem-2shapes}.
According to (1) and (4) of Lemma~\ref{lem-2shapes}, $l$ has two $I$-vanishing sides (say $H_1$ and $H_2$), and the two arcs in $\partial I$ connecting $u,v$ are contained in $\partial C_{p_1}$ and $\partial C_{p_2}$ respectively (we denote by $\sigma_i$ the arc contained in $C_{p_i}$ for $i \in \{1,2\}$).
Without loss of generality, assume $\sigma_i \subseteq H_i$ for $i \in \{1,2\}$.

%By construction, either $I \cap V = C_{p_1} \cap V$ or $I \cap V = C_{p_2} \cap V$ (assume the former without loss of generality).
First, we prove the connectedness of $\partial C_p \cap I$.
We may assume $H_1$ is $C_p$-vanishing (as the ``roles'' of $H_1$ and $H_2$ are symmetric).
Since $p,p_1,p_2$ are not collinear, by Fact~\ref{fact-degen} we have $I \cap l \neq C_p \cap l$.
So we consider two cases: $C_p \cap l \subsetneq I \cap l$ and $C_p \cap l \nsubseteq I \cap l$.
%If $C_p \cap l = I \cap l$, then by Fact~\ref{fact-degen} $p,p_1,p_2$ are collinear, contradicting the assumption.
If $C_p \cap l \subsetneq I \cap l$, then $u,v \notin C_p^\circ$.
By applying Lemma~\ref{lem-exclude} (recall that we already assumed $I^\circ \cap C_p^\circ \neq \emptyset$), we have $C_p \cap I = C_p \cap C_{p_i}$ for some $i \in \{1,2\}$.
Therefore, $\partial C_p \cap I = \partial C_p \cap C_{p_i}$, which is connected by Lemma~\ref{lem-2shapes}.
If $C_p \cap l \nsubseteq I \cap l$, then there must be an endpoint $e$ of $C_p \cap l$ that is not in $I$.
We have $\partial C_p \cap I = (\partial C_p \backslash \{e\}) \cap I = L_1 \cap L_2$, where $L_i = (\partial C_p \backslash \{e\}) \cap C_{p_i}$ for $i = \{1,2\}$.
Note that $\partial C_p \backslash \{e\}$ is homeomorphic to the real line $\mathbb{R}$ and the intersection of any connected subsets of $\mathbb{R}$ is connected.
Therefore, it suffices to show that $L_1$ and $L_2$ are connected subsets of $\partial C_p \backslash \{e\}$.
If $e \notin C_{p_1}$, then $L_1 = \partial C_p \cap C_{p_1}$, whose connectedness is implied by Lemma~\ref{lem-2shapes}.
If $e \in C_{p_1}$, then $e \in C_{p_1} \cap H_1$.
But $e \notin I \cap H_1$, as $e \notin I \cap l$.
It follows that $C_{p_1} \cap H_1 \neq I \cap H_1$.
This happens only when $\sigma_1 \subseteq l$, i.e., $\sigma_1$ is the segment $[u,v]$.
In this situation, $C_{p_1} = C_{p_1} \cap H_2$ and hence $L_1 = (\partial C_p \cap C_{p_1} \cap H_2) \backslash \{e\}$.
Let $K = \partial C_p \cap H_2$, then $L_1 = (K \cap C_{p_1}) \cap (K \backslash \{e\})$.
Now $K \cap C_{p_1} = \partial C_p \cap C_{p_1} \cap H_2 = \partial C_p \cap C_{p_1}$ is connected by Lemma~\ref{lem-2shapes} and $K \backslash \{e\}$ is also connected (as $K$ is indeed a simple curve and $e$ is an endpoint of $K$).
Since $K$ is homeomorphic to the closed interval $[0,1]$ and the intersection of any connected subsets of $[0,1]$ is connected, we deduce that $L_1$ is connected.
Using the same argument, we can prove the connectedness of $L_2$.
Consequently, $\partial C_p \cap I = L_1 \cap L_2$ is connected.

\begin{figure}[htbp]
    \begin{center}
        \includegraphics[height=5cm]{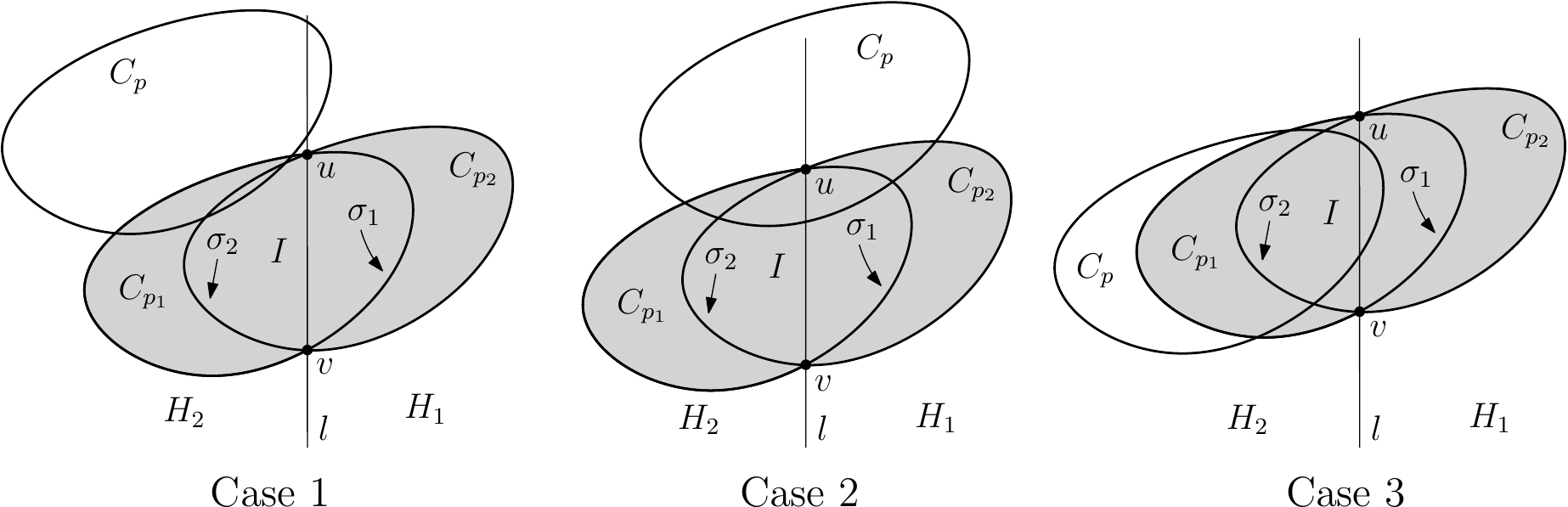}
    \end{center}
    \caption{Illustrating the three cases}
    \label{fig-prf7}
\end{figure}

Next, we prove the connectedness of $\partial I \cap C_p$.
%Let $e_1,e_2$ be the two endpoints of $I \cap l$.
If $I \cap l \subseteq C_p$, then by Lemma~\ref{lem-contain} we have $I \subseteq C_p$ and thus $\partial I \cap C_p = \partial I$ is connected.
If $I \cap l \nsubseteq C_p$, we investigate three cases.
In the first case, $I \cap l$ and $C_p$ are disjoint; see the left figure of Figure~\ref{fig-prf7}.
Then we have either $C_p \cap I \subseteq (H_1 \backslash l)$ or $C_p \cap I \subseteq (H_2 \backslash l)$, as $C_p \cap I$ is connected.
Therefore, either $\sigma_1 \cap C_p = \emptyset$ or $\sigma_2 \cap C_p = \emptyset$.
%We claim that either $\sigma_1 \cap C_p = \emptyset$ or $\sigma_2 \cap C_p = \emptyset$.
%If $\sigma_i \subseteq l$ for some $i \in \{1,2\}$, we are done, because $\sigma_i \cap C_p = I \cap l \cap C_p = \emptyset$.
%Otherwise, we have $I \cap H_i = C_{p_i} \cap H_i = \mathcal{CH}(\sigma_i)$ (and in particular $I \cap l = C_{p_i} \cap l$) for $i \in \{1,2\}$.
%As such, $C_p \cap C_{p_1} \cap l = C_p \cap I \cap l = \emptyset$.
%This implies either $C_p \cap C_{p_1} \cap H_1 = \emptyset$ or $C_p \cap C_{p_1} \cap H_2 = \emptyset$, since $C_p \cap C_{p_1}$ is connected.
%If $C_p \cap C_{p_1} \cap H_1 = \emptyset$, then $C_p \cap I \cap H_1 = \emptyset$.
%In this situation, we have $\partial I \cap C_p = \partial I \cap H_2 \cap C_p = \sigma_2 \cap C_p$.
%It follows that $(\partial I \backslash \sigma_2) \cap C_p = \emptyset$.
%Since $u,v \notin C_p$, we have $\sigma_1 \cap C_p = (\{u,v\} \cup (\partial I \backslash \sigma_2)) \cap C_p = \emptyset$.
%Similarly, if $C_p \cap C_{p_1} \cap H_2 = \emptyset$, then $C_p \cap I \cap H_2 = \emptyset$, and in this situation we can deduce $\partial I \cap C_p = \partial I \cap H_1 \cap C_p = \sigma_1 \cap C_p$.
%Using the same argument as above, we have $\sigma_2 \cap C_p = \emptyset$.
%Based on this observation, we can show the connectedness of $\partial I \cap C_p$ as follows.
Without loss of generality, we assume $\sigma_1 \cap C_p = \emptyset$.
Then $\partial I \cap C_p = \sigma_2 \cap C_p$.
Note that $\sigma_2$ is a connected portion of $\partial C_{p_2}$.
Since $u,v \notin C_p$ and $\partial C_{p_2} \cap C_p$ is connected by Lemma~\ref{lem-2shapes}, we have either $\partial C_{p_2} \cap C_p \subseteq \sigma_2$ (then $\sigma_2 \cap C_p = \partial C_{p_2} \cap C_p$) or $\partial C_{p_2} \cap C_p \subseteq \partial C_{p_2} \backslash \sigma_2$ (then $\sigma_2 \cap C_p = \emptyset$).
It follows that $\sigma_2 \cap C_p$ is connected, and so is $\partial I \cap C_p$.
This completes the first case.
Then we study the second case, in which $C_p$ contains one of $u,v$ but does not contain the other one (assume $u \in C_p$ and $v \notin C_p$); see the middle figure of Figure~\ref{fig-prf7}.
In this case, we claim that $\sigma_i \cap C_p$ is connected for all $i \in \{1,2\}$.
It suffices to show the connectedness of $\sigma_1 \cap C_p$.
Since $v \notin C_p$ and $\sigma_1 \subseteq \partial C_{p_1}$, we have $\sigma_1 \cap C_p = (\sigma_1 \backslash \{v\}) \cap (\partial C_{p_1} \cap C_p)$.
Now $\sigma_1 \backslash \{v\}$ and $\partial C_{p_1} \cap C_p$ are both connected subsets of $\partial C_{p_1} \backslash \{v\}$ (the connectedness of the latter is implied by Lemma~\ref{lem-2shapes}), and $\partial C_{p_1} \backslash \{v\}$ is homeomorphic to the real line $\mathbb{R}$.
Therefore, $\sigma_1 \cap C_p$ is connected.
Symmetrically, $\sigma_2 \cap C_p$ is also connected.
Note that both $\sigma_1 \cap C_p$ and $\sigma_2 \cap C_p$ contain the point $u$, hence $(\sigma_1 \cap C_p) \cup (\sigma_2 \cap C_p) = \partial I \cap C_p$ is connected.
This completes the second case.
Finally, we consider the last case, in which $C_p \cap l \subseteq (I \cap l) \backslash \{u,v\}$; see the right figure of Figure~\ref{fig-prf7}.
Without loss of generality, assume $H_1$ is $C_p$-vanishing.
We claim that $\sigma_1 \cap C_p = \emptyset$.
%If $\sigma_1 \nsubseteq l$, then $\sigma_1 = \partial C_{p_1} \cap H_1$.
Since $C_p \cap l \subseteq (I \cap l) \backslash \{u,v\} \subseteq I \cap l \subseteq C_{p_1} \cap l$, we can apply (2) of Corollary~\ref{cor-halfcontain} to deduce that $C_p \cap (H_1 \backslash l) \subseteq C_{p_1} \backslash \partial C_{p_1}$, which implies $\sigma_1 \cap C_p \subseteq l$.
If $\sigma_1 \nsubseteq l$, then $\sigma_1 \cap l = \{u,v\}$.
Since $u,v \notin C_p$, we have $\sigma_1 \cap C_p = \sigma_1 \cap C_p \cap l = \emptyset$.
If $\sigma_1 \subseteq l$, then $\sigma_1$ is the segment $[u,v]$.
In this situation, we must have $C_{p_1} \subseteq H_2$.
Without loss of generality, we assume that $l$ is the line $x=0$ and $H_1$ is the side $x \leq 0$.
Because $\text{len}_{C_p}(l) < \text{len}_{C_{p_1}}(l)$ (as $C_p \cap l \subsetneq C_{p_1} \cap l$) and $H_1$ is $C_p$-vanishing by assumption, we have $p.x > p_1.x$.
This implies $C_p \subseteq H_2 \backslash l$ (as $C_{p_1} \subseteq H_2$) and hence $\sigma_1 \cap C_p = \emptyset$.
Therefore, we have $\partial I \cap C_p = (\sigma_1 \cup \sigma_2) \cap C_p = \sigma_2 \cap C_p$.
But $\sigma_2$ is a connected portion of $\partial C_{p_2}$.
Since $u,v \notin C_p$ and $\partial C_{p_2} \cap C_p$ is connected by Lemma~\ref{lem-2shapes}, we have either $\partial C_{p_2} \cap C_p \subseteq \sigma_2$ (then $\sigma_2 \cap C_p = \partial C_{p_2} \cap C_p$) or $\partial C_{p_2} \cap C_p \subseteq \partial C_{p_2} \backslash \sigma_2$ (then $\sigma_2 \cap C_p = \emptyset$).
It follows that $\sigma_2 \cap C_p$ is connected, and so is $\partial I \cap C_p$.
This completes the last case as well as the entire proof.

\ifdefined\SODA
\vspace{-0.25cm}
\fi
\subsection{Proof of Lemma~\ref{lem-interior}}
\ifdefined\SODA
\vspace{-0.15cm}
\fi
We first establish an easy fact.
\ifdefined\SODA
\vspace{-0.1cm}
\fi
\begin{fact} \label{fact-plnint}
    If two convex bodies $C$ and $D$ in $\mathbb{R}^2$ plainly intersect, then $\partial C \cap D^\circ$ and $\partial D \cap C^\circ$ are both connected.
\end{fact}
\ifdefined\SODA
\vspace{-0.1cm}
\fi
\textit{Proof.}
Clearly, we have $\partial C \cap D \subseteq \partial I$ and $\partial I = (\partial C \cap D) \cup (\partial D \cap C)$.
It follows that $\partial C \cap D^\circ = \partial I \backslash \partial D = \partial I \backslash (\partial D \cap C)$.
Since $\partial D \cap C$ is connected, $\partial C \cap D^\circ$ is connected.
For the same reason, $\partial D \cap C^\circ$ is connected.
\hfill $\Box$
\smallskip
\begin{figure}[htbp]
    \begin{center}
        \includegraphics[height=4.5cm]{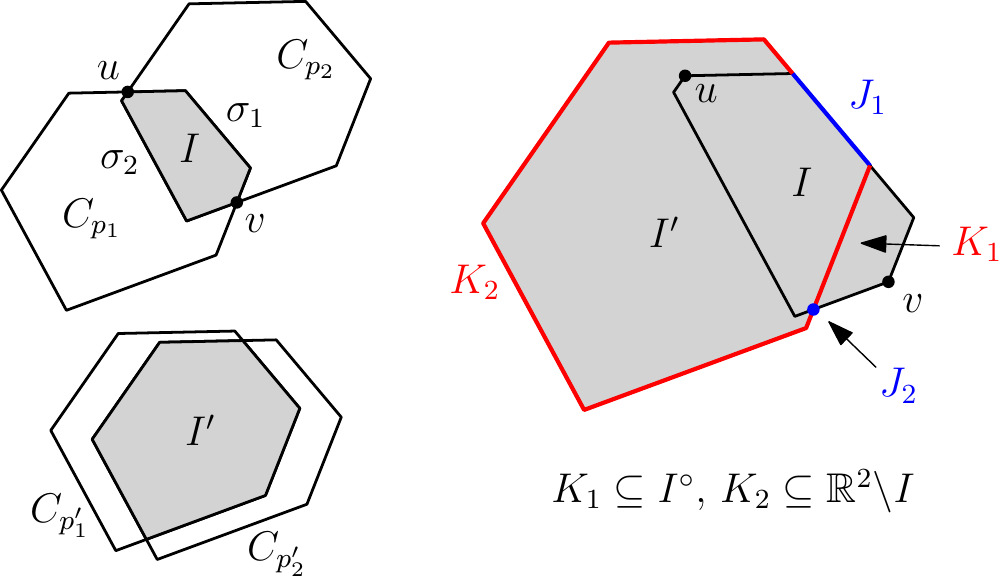}
    \end{center}
    \caption{Illustrating the case that $K_2 \subseteq \mathbb{R}^2 \backslash I$}
    \label{fig-prf8}
\end{figure}

\noindent
Now we are able to prove the lemma.
Assume $u \in I'^\circ$.
Let $\sigma_1,\sigma_2$ be the two arcs in $\partial I$ connecting $u,v$.
By (4) of Lemma~\ref{lem-2shapes}, we may assume $\sigma_i \subseteq \partial C_{p_i}$ for $i \in \{1,2\}$.
We consider three cases: $v \in I'^\circ$, $v \in \partial I'$, $v \notin I'$.
If $v \in I'^\circ$, then $v \in C_{p_i'}^\circ$ for all $i \in \{1,2\}$.
By Lemma~\ref{lem-contain}, we have $I \subseteq C_{p_i'}^\circ$ for $i \in \{1,2\}$ and thus $I \subseteq I'^\circ$.
Therefore, $\partial I \cap I' = \partial I$ and $\partial I' \cap I = \emptyset$, which implies that $I$ and $I'$ plainly intersect.
If $v \in \partial I'$, we can apply Lemma~\ref{lem-contain} to deduce $I \subseteq C_{p_i'}$ for $i \in \{1,2\}$.
Hence, $I \subseteq I'$ and $\partial I \cap I' = \partial I$.
It suffices to show the connectedness of $\partial I' \cap I$.
Since $I^\circ \subseteq I'^\circ$, we have $\partial I' \cap I^\circ = \emptyset$ and thus $\partial I' \cap I = \partial I' \cap \partial I = (\partial I' \cap \sigma_1) \cup (\partial I' \cap \sigma_2)$.
We claim that $\partial I' \cap \sigma_i$ is connected for all $i \in \{1,2\}$.
%Note that $\partial I' \cap \sigma_1 = (\sigma_1 \cap I') \backslash (\sigma_1 \cap I'^\circ) = \sigma_1 \backslash (\sigma_1 \cap I'^\circ)$.
Since $v \notin I'^\circ$ and $\sigma_1 \subseteq \partial C_{p_1}$, we have $\sigma_1 \cap I'^\circ = (\partial C_{p_1} \cap I'^\circ) \cap (\sigma_1 \backslash \{v\})$.
Note that both $\partial C_{p_1} \cap I'^\circ$ and $\sigma_1 \backslash \{v\}$ are connected subsets of $\partial C_{p_1} \backslash \{v\}$ (the connectedness of the former is implied by Lemma~\ref{lem-3shapes} and Fact~\ref{fact-plnint}).
But $\partial C_{p_1} \backslash \{v\}$ is homeomorphic to the real line $\mathbb{R}$ and the intersection of any connected subsets of $\mathbb{R}$ is connected.
Therefore, $\sigma_1 \cap I'^\circ$ is a connected portion of $\sigma_1$.
It follows that $\sigma_1 \backslash (\sigma_1 \cap I'^\circ)$ is connected, because $u \in \sigma_1 \cap I'^\circ$ and $u$ is an endpoint of $\sigma_1$.
The connectedness of $\partial I' \cap \sigma_1$ is then implied by the fact that $\partial I' \cap \sigma_1 = \sigma_1 \backslash (\sigma_1 \cap I'^\circ)$.
Symmetrically, we can show that $\partial I' \cap \sigma_2$ is also connected.
Applying the fact that both $\partial I' \cap \sigma_1$ and $\partial I' \cap \sigma_2$ contain $u$, we finally deduce that $\partial I' \cap \partial I = (\partial I' \cap \sigma_1) \cup (\partial I' \cap \sigma_2)$ is connected.
The rest of the proof is dedicated to the case $v \notin I'$.
In this case, we again claim that $\partial I' \cap \sigma_i$ is connected for all $i \in \{1,2\}$.
Using the same argument as above, we can deduce that $\sigma_1 \cap I'^\circ$ is connected (note that when proving the connectedness of $\sigma_1 \cap I'^\circ$ in the previous case, we only used the fact $v \notin I'^\circ$).
Also, a similar argument applies to show the connectedness of $\sigma_1 \cap I'$.
Indeed, $\sigma_1 \cap I' = (\sigma_1 \backslash \{v\}) \cap (\partial C_{p_1} \cap I')$.
Now both $\sigma_1 \backslash \{v\}$ and $\partial C_{p_1} \cap I'$ are connected subsets of $\partial C_{p_1} \backslash \{v\}$, and $\partial C_{p_1} \backslash \{v\}$ is homeomorphic to $\mathbb{R}$.
This implies the connectedness of $\sigma_1 \cap I'$.
Since both $\sigma_1 \cap I'^\circ$ and $\sigma_1 \cap I'$ are connected portions of $\sigma_1$ containing $u$, we deduce that $\partial I' \cap \sigma_1 = (\sigma_1 \cap I') \backslash (\sigma_1 \cap I'^\circ)$ is connected.
Symmetrically, $\partial I' \cap \sigma_2$ is also connected.
Note that $u,v \notin \partial I' \cap \sigma_i$ for all $i \in \{1,2\}$.
Therefore, $\partial I' \cap \partial I$ consists of two connected components, namely, $J_1 = \partial I' \cap \sigma_1$ and $J_2 = \partial I' \cap \sigma_2$.
It follows that $\partial I' \backslash \partial I = \partial I' \backslash (J_1 \cap J_2)$ also consists of two connected components, say $K_1$ and $K_2$.
By the connectedness of each $K_i$, we have either $K_i \subseteq I^\circ$ or $K_i \subseteq \mathbb{R}^2 \backslash I$.
If $K_i \subseteq \mathbb{R}^2 \backslash I$ for all $i \in \{1,2\}$, then either $I^\circ \subseteq I'$ or $I^\circ \cap I' = \emptyset$.
The former is not true as $v \notin I'$ (one can take a point $r$ sufficiently close to $v$ such that $r \in I^\circ$ and $r \notin I'$), while the latter is not true as $u \in I'^\circ$ (one can take a point $r$ sufficiently close to $u$ such that $r \in I^\circ$ and $r \in I'^\circ$).
As such, $K_i \subseteq I^\circ$ for some $i \in \{1,2\}$, say $K_1 \subseteq I^\circ$.
In this situation, either $\partial I' \cap I = \partial I'$ (if $K_2 \subseteq I^\circ$) or $\partial I' \cap I = J_1 \cup J_2 \cup K_1 = \partial I' \backslash K_2$ (if $K_2 \subseteq \mathbb{R}^2 \backslash I$), hence $\partial I' \cap I$ is connected; see Figure~\ref{fig-prf8} for an illustration of the latter case. 
To show the connectedness of $\partial I \cap I'$ is easier.
In the above argument, we already show that $\sigma_i \cap I'$ is connected for all $i \in \{1,2\}$.
Because both $\sigma_1 \cap I'$ and $\sigma_2 \cap I'$ contain $u$, it follows that $\partial I \cap I' = (\sigma_1 \cap I') \cup (\sigma_2 \cap I')$ is connected.

\ifdefined\SODA
\vspace{-0.25cm}
\fi
\subsection{Proof of Lemma~\ref{lem-topbottom}}
\begin{figure}[htbp]
    \begin{center}
        \includegraphics[height=4.5cm]{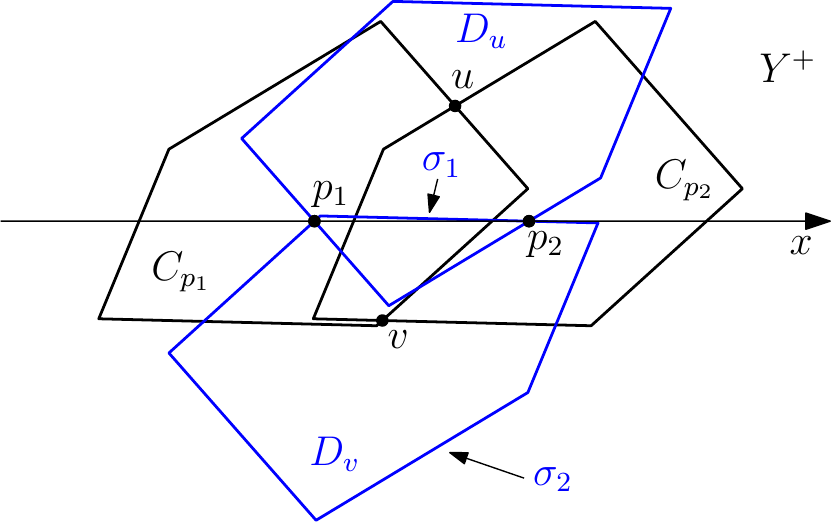}
    \end{center}
    \caption{Illustrating the proof of Lemma~\ref{lem-topbottom}}
    \label{fig-prf9}
\end{figure}

\ifdefined\SODA
\vspace{-0.15cm}
\fi
We define $D = \{(x,y) \in \mathbb{R}^2: (-x,-y) \in C\}$.
Intuitively, $D$ is the convex body in $\mathbb{R}^2$ obtained by rotating $C$ around the origin by an angle of $\pi$ (either clockwise or counterclockwise).
One can easily verify that $u \in C_p$ (resp., $v \in C_p$) iff $p \in D_u$ (resp., $p \in D_v$).
Therefore, to show that $v \in C_p$ only if $u \in C_p$ for any point $p \in \mathbb{R}^2$ with $p.y>0$, it suffices to show that $D_v \cap Y^+ \subseteq D_u$, where $Y^+ = \{(x,y) \in \mathbb{R}^2: y>0\}$.
Since $u \in \partial C_{p_i}$ for $i \in \{1,2\}$, we have $p_i \in \partial D_u$ for $i \in \{1,2\}$.
For the same reason, $p_i \in \partial D_v$ for $i \in \{1,2\}$.
Consider the two arcs $\sigma_1,\sigma_2$ in $\partial D_v$ connecting $p_1,p_2$.
Clearly, one arc is on the top side of the $x$-axis (say $\sigma_1$), while the other one is on the bottom side (say $\sigma_2$).
See Figure~\ref{fig-prf9} for an illustration of the notations.
%If both $\sigma_1,\sigma_2$ are on the bottom side of the $x$-axis, then $D_v \cap Y^+ = \emptyset$ and we are done.
%Otherwise, suppose $\sigma_1 \cap Y^+ \neq \emptyset$.
We claim that $\sigma_1 \subseteq D_u$.
Note that $D_u$ and $D_v$ plainly intersect by Lemma~\ref{lem-2shapes}, hence $\partial D_v \cap D_u$ is connected.
Since $p_1,p_2 \in \partial D_v \cap D_u$, either $\sigma_1 \subseteq D_u$ or $\sigma_2 \subseteq D_u$.
So it suffices to show that $\sigma_2 \nsubseteq D_u$.
Let $s$ be a bottommost point of $D_v$, i.e., a point with a minimum ordinate.
Then $s \in \partial D_v$.
Because $v.y<u.y$, $s$ has a smaller ordinate than any point in $D_u$.
In particular, $s \notin D_u$.
Furthermore, we must have $s.y<0$ (as $p_1.y = p_2.y = 0$ and $p_1,p_2 \in D_u$), which implies $s \in \sigma_2$.
It follows that $\sigma_2 \nsubseteq D_u$ and thus $\sigma_1 \subseteq D_u$.
Therefore, $D_v \cap Y^+ = \mathcal{CH}(\sigma_1) \cap Y^+ \subseteq D_u$, as desired.
Next, we show that $v \in C_p^\circ$ only if $u \in C_p^\circ$ for any point $p \in \mathbb{R}^2$ with $p.y>0$.
Assume $v \in C_p^\circ$ and $u \notin C_p^\circ$.
Since $v \in C_p^\circ \subseteq C_p$, by the previous proof, we can deduce that $u \in C_p$ and thus $u \in \partial C_p$.
Now one can take a point $p' \in Y^+$ sufficiently close to $p$ such that $v \in C_{p'}^\circ$ and $u \notin C_{p'}$.
But by the previous proof, we must have $v \in C_{p'}$ only if $u \in C_{p'}$.
This contradiction completes the proof.

\section{Finding the nonempty cells intersecting a $\varGamma$-translate} \label{appx-cell}
In this section, we show how to build an efficient data structure to report the nonempty cells (and quad-cells) intersecting a given $\varGamma$-translate.
%We only need to consider cells, as the quad-cells can be found once the cells are found.

Let $\overline{G} = \{\Box \in G : \Box \cap S \neq \emptyset\}$ denote the set of all nonempty cells of $G$.
Assume that the grid length is $\ell$ and the diameter of $\varGamma$ is $\Delta$.
Note that both $\ell$ and $\Delta$ are constants as they only depend on $\varGamma$ and $\varGamma$ is fixed.
Now let $L$ be a sufficiently large constant such that $L > 2(\Delta+\ell)$.

For each $\Box \in \overline{G}$, let $c_\Box$ be the center point of $\Box$ and $\text{Cand}(\Box) = \{\Box' \in \overline{G} : \text{dist}(c_\Box,c_{\Box'}) \leq L\}$.
We have $|\text{Cand}(\Box)| = O(L^2 / \ell^2) = O(1)$.
Next, we build a Voronoi Diagram $\mathcal{VD}$ on the point-set $A = \{c_\Box: \Box \in \overline{G}\}$ and associate to each site $c_\Box$ the set $\text{Cand}(\Box)$.
Clearly, $\mathcal{VD}$ occupies $O(n)$ space since $|\overline{G}| = O(n)$ and $|\text{Cand}(\Box)| = O(1)$ for all $\Box \in \overline{G}$.

Fix a point $o \in \varGamma$.
Given a query $\varGamma_q \in \mathcal{L}_\varGamma$, we query $\mathcal{VD}$ with the point $o+q$ to obtain the nearest-neighbor of $o+q$ in $A$, say $c_{\Box^*}$, in $O(\log n)$ time.
We claim that all the nonempty cells intersecting $\varGamma_q$ are contained in $\text{Cand}(\Box^*)$.
Let $\Box \in \overline{G}$ be a cell intersecting $\varGamma_q$.
Since $o \in \varGamma$, we have $o+q \in \varGamma_q$.
Therefore, $\text{dist}(o+q,c_\Box) \leq D+\ell$.
Since $c_{\Box^*}$ is the nearest-neighbor of $o+q$ in $A$ and $c_\Box \in A$, we have $\text{dist}(o+q,c_{\Box}^*) \leq \text{dist}(o+q,c_\Box) \leq D+\ell$.
By the triangle inequality, $\text{dist}(c_\Box,c_{\Box}^*) \leq 2(\Delta+\ell) < L$.
Thus, $\Box \in \text{Cand}(\Box^*)$ and our claim holds.
Now we only need to check the $O(1)$ cells in $\text{Cand}(\Box^*)$ one-by-one, and all the nonempty cells intersecting $\varGamma_q$ can be found.

Reporting the nonempty quad-cells intersecting $\varGamma_q$ can be simply done once the nonempty cells intersecting $\varGamma_q$ are found.

\section{Reporting the shortest pair in a co-wedge translate} \label{appx-cowedge}
%把3写到这个section，愿付公与你同在！
%好的，我估计今晚写好，到时候让付公给你看一下。
\begin{figure}[t]
    \centering
    \includegraphics[height=2.7cm]{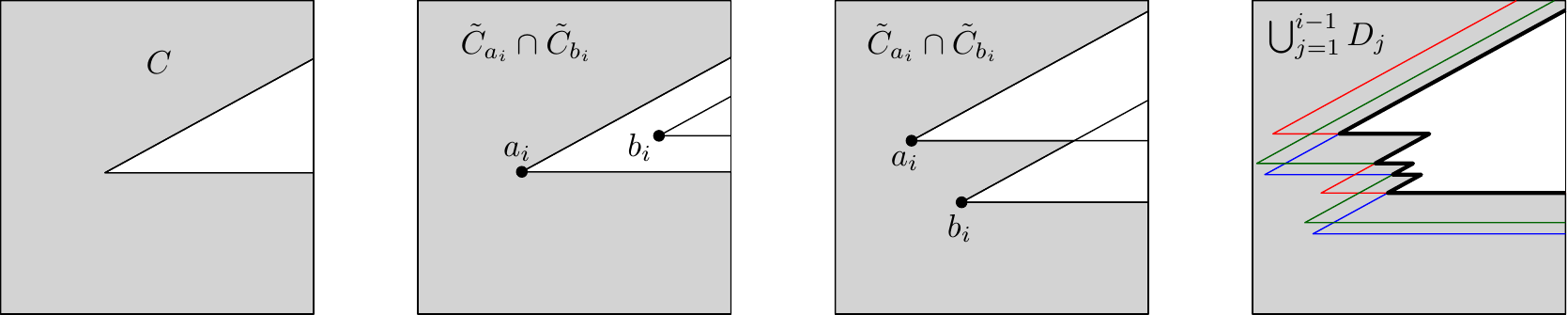}
    \caption{Illustrating a co-wedge, two examples of $\tilde{C}_{a_i} \cap \tilde{C}_{b_i}$, and an example of $\bigcup_{j=1}^{i-1} D_j$.}
    \label{figure:co_wedge}
\end{figure}
\noindent
Suppose $\varPhi(S,\mathcal{L}_C) = \{\phi_1,\dots,\phi_m\}$ where $\phi_i = (a_i,b_i)$ and $\phi_1,\dots,\phi_m$ are sorted in increasing order of their lengths.
We want to build a data structure which can report, for a query $C_q \in \mathcal{L}_C$, the smallest $i$ such that $a_i,b_i \in C_q$.

%We have $m = O(n)$ by Lemma~\ref{lem-linearwedcand}.
The idea is similar to that in Section~\ref{Handling wedge translation queries} for wedges but involves a more careful argument.
%Again, suppose $\varPhi(S,\mathcal{L}_C) = \{\phi_1,\dots,\phi_m\}$ where $\phi_i = (a_i,b_i)$ and $\phi_1,\dots,\phi_m$ are sorted in increasing order of their lengths, where $m = O(n)$ according to Lemma~\ref{lem-linearcowcand}. 
Let $\tilde{C} = \{(x,y):(-x,-y) \in C\}$, which is a co-wedge obtained by rotating $C$ around the origin with angle $\pi$.
For a point $p \in \mathbb{R}^2$, it is clear that $a_i,b_i \in C_p$ iff $p \in \tilde{C}_{a_i} \cap \tilde{C}_{b_i}$.
Unlike the wedge case, depending on the position of $a_i$ and $b_i$, $\tilde{C}_{a_i} \cap \tilde{C}_{b_i}$ may be a polygonal region whose boundary consists of at most four edges and three vertices; see the middle two figures in Figure~\ref{figure:co_wedge} for an example.
Denote $D_i = \tilde{C}_{a_i} \cap \tilde{C}_{b_i}$. 
By successively overlaying $D_1, \dots, D_m$, we obtain a planar subdivision whose cells are $\varSigma_1,\dots,\varSigma_m$ where $\varSigma_i = D_i \backslash \bigcup_{j=1}^{i-1} D_j$.
(Note that, unlike the wedge-case, $\varSigma_i$ here might not necessarily be connected.)
Then, the answer of a query $C_q \in \varGamma_C$ is $i$ iff $q \in \varSigma_i$.

It now suffices to analyze the complexity of the subdivision and then build on it an optimal point-location data structure to answer each query efficiently.
As the subdivision forms a planar graph, it suffices to bound the total number of vertices.
We argue that at most $O(1)$ new vertices can be created after we overlay $D_i$.
%that is, besides the two vertices $a_i$ and $b_i$, the boundary of $D_i$ can only intersect with the boundary of $\bigcup_{j=1}^{i-1} D_j$ constant times.
To see why, we note that the boundary of $\bigcup_{j=1}^{i-1} D_j$ consists of segments that are parallel to one branch of $\tilde{C}$.
See the rightmost figure in Figure~\ref{figure:co_wedge} for an example.
Furthermore, any line parallel to one branch of $C$ intersects the boundary of $\bigcup_{j=1}^{i-1} D_j$ at most once.
The boundary of $D_i$ consists of at most four edges, each of which is a segment or a ray parallel to one branch of $\tilde{C}$.
As such, the boundary of $D_i$ can intersect the boundary of $\bigcup_{j=1}^{i-1} D_j$ at most four times.
Including the vertices (at most three) of $D_i$, at most seven new vertices are created after we overlay $D_i$.
%This means that we can partition $\partial \bigcup_{j=1}^{i-1} D_j$ into two disjoint sets, each of which contains only parallel segments.
%It then follows that every edge of $D_i$ can intersect with at most one segment in either set. This is because if the edge intersects with (say) two parallel segments at two different intersections, then the farther intersection can never be on the boundary of $\bigcup_{j=1}^{i-1} D_j$ --- a contradiction.
Therefore, after overlaying all $D_1,\dots,D_m$, the complexity of the eventual subdivision is $O(m)$.
As such, the data structure uses $O(m)$ space and $O(\log m)$ query time.

\end{document}